\documentclass[aps,pra,longbibliography,reprint,showpacs,superscriptaddress,footinbib,citeautoscript]{revtex4-1}

\usepackage{graphicx}
\usepackage{amsmath}
\usepackage{amssymb}
\usepackage{mathrsfs}
\usepackage{bm}
\usepackage{array}
\newcolumntype {s}[1]{@{\hspace{#1}}} 
\usepackage[usenames,dvipsnames]{xcolor}
\usepackage{wrapfig}
\usepackage[normalem]{ulem}

\usepackage[colorlinks=true,citecolor=cyan,linkcolor=magenta]{hyperref}

\graphicspath{{.}{./FinalFigures/}{./EPS/}}

\newcommand* {\ee}{\mathrm{e}}

\newcommand*{\vek}[1]{{\bm{\mathrm{#1}}}}

\newcommand*{\kk}{{\bm{\mathrm{k}}}}
\newcommand*{\ssig}{{\bm{\mathrm{\sigma}}}}

\newcommand*{\qq}{{\bm{\mathrm{q}}}}

\newcount\colveccount
\newcommand*\colvec[1]{
        \global\colveccount#1
        \begin{pmatrix}
        \colvecnext
}
\def\colvecnext#1{
        #1
        \global\advance\colveccount-1
        \ifnum\colveccount>0
                \\
                \expandafter\colvecnext
        \else
                \end{pmatrix}
        \fi
}

\def\lsim{\raise0.3ex\hbox{$\;<$\kern-0.75em\raise-1.1ex\hbox{$\sim\;$}}}
\def\gsim{\raise0.3ex\hbox{$\;>$\kern-0.75em\raise-1.1ex\hbox{$\sim\;$}}}

\DeclareMathSymbol{\myRe}{\mathord}{symbols}{"3C}

\DeclareMathSymbol{\myIm}{\mathord}{symbols}{"3D}

\begin{document}

\title{Anomalous Spin Response and Virtual-Carrier-Mediated Magnetism in a Topological
Insulator}

\author{T. Kernreiter}
\affiliation{School of Chemical and Physical Sciences and MacDiarmid Institute
for Advanced Materials and Nanotechnology, Victoria University of Wellington,
PO Box 600, Wellington 6140, New Zealand}

\author{M. Governale}
\email{michele.governale@vuw.ac.nz}
\affiliation{School of Chemical and Physical Sciences and MacDiarmid Institute
for Advanced Materials and Nanotechnology, Victoria University of Wellington,
PO Box 600, Wellington 6140, New Zealand}

\author{U. Z\"ulicke}
\email{uli.zuelicke@vuw.ac.nz}
\affiliation{School of Chemical and Physical Sciences and MacDiarmid Institute
for Advanced Materials and Nanotechnology, Victoria University of Wellington,
PO Box 600, Wellington 6140, New Zealand}

\author{E. M. Hankiewicz}
\affiliation{Institut f\"ur Theoretische Physik und Astrophysik (TP4),
Universit\"at W\"urzburg, 97074 W\"urzburg, Germany}

\date{\today}

\begin{abstract}

We present a comprehensive theoretical study of the static spin response in HgTe
quantum wells, revealing distinctive behavior for the topologically nontrivial inverted
structure. Most strikingly, the $\qq=0$ (long-wave-length) spin susceptibility of the
undoped topological-insulator system is constant and equal to the value found
for the gapless Dirac-like structure, whereas the same quantity shows the typical
decrease with increasing band gap in the normal-insulator regime. We discuss
ramifications for the ordering of localized magnetic moments present in the quantum
well, both in the insulating and electron-doped situations. The spin response of edge
states is also considered, and we extract effective Land\'e $g$-factors for the bulk
and edge electrons. The variety of counter-intuitive spin-response properties revealed in
our study arises from the system's versatility in accessing situations where the charge-carrier
dynamics can be governed by ordinary Schr\"odinger-type physics, mimics the behavior of chiral
Dirac fermions, or reflects the material's symmetry-protected topological order.

\end{abstract}


\maketitle

\section{Introduction \& Main Results}

The two-dimensional (2D) topological insulator (TI) realized in inverted HgTe quantum wells exhibits
unusual electric-transport properties~\cite{Koenig2007Scien,Roth2009Scien,Brune2012NatPhys,
Hart2015arXiv} that have attracted great interest~\cite{Qi2011RMP}. TI behavior is also found in other
2D~\cite{Liu2008PRLa,Knez2011PRL} and bulk~\cite{Qi2011RMP,Hasan2011AnnuRev} materials.
The potential for interesting interplays between a TI's electronic and magnetic degrees of freedom,
e.g., through hyperfine interaction with the material's nuclei~\cite{Lunde2013PRB}, or local exchange
interaction with magnetic dopants~\cite{Novik2005PRB}, has been pointed out
recently~\cite{Liu2008PRLb,Yu2010Scien}. These efforts have opened up new perspectives and
extended previous work devoted to understanding spin effects~\cite{Simon2007PRL,Simon2008PRB}
and magnetism~\cite{Dietl2010NatMat} in ordinary semiconductors. Our present theoretical study of
the spin response in HgTe quantum wells reveals unconventional spin-related properties that
distinguish this paradigmatic TI material from all other currently known 2D electronic systems. We
thus provide alternative means for the experimental identification of the topological regime and extend
current knowledge about the fundamentals of spin-response behavior in solids.

The spin susceptibility contains comprehensive information about the magnetic properties of a
material. In the simplest case of a spin-rotationally invariant noninteracting electron gas, the spin
susceptibility is proportional to the charge-response (Lindhard) function~\cite{vignalebook}.
Noticable deviations from that situation occur, e.g., in systems with strong spin-orbit coupling
such as 2D hole gases~\cite{Kernreiter2013PRL}. For metals or degenerately doped semiconductors,
the spin response of only the partially filled band is typically considered. This approach misses
intrinsic contributions to many-particle response functions arising from virtual interband transitions
that become important in narrow-gap and, especially gapless, electron systems. Examples for the
latter are the 2D Dirac-like electron states on the surface of a bulk TI whose magnetic properties
have been discussed in Refs.~\onlinecite{Liu2009PRL,Zhu2011PRL}. The HgTe quantum wells
considered here present an ideal testing ground for exploring the importance and properties of
intrinsic, or virtual-carrier, effects, as it is possible to tune the band gap in such systems with a
single structural parameter (i.e., the quantum-well width $d$). Furthermore, deviations from the
conical 2D-Dirac dispersion in the gapless case are well characterized within a continuum-model
(BHZ) description~\cite{Bernevig2006} that also gives controlled access to the full range of, and
interesting transitions between, Schr\"odinger-physics-dominated, chiral-Dirac-fermion-like, and
topologically nontrivial phases. In particular, definite (i.e., cut-off-independent) results for the intrinsic
response at long wavelengths are obtained even in the limit of vanishing band gap; unlike in the case
of the previously considered 2D-Dirac models used to describe the surface states of bulk
TIs~\cite{Liu2009PRL,Zhu2011PRL}.

Before going into greater detail in the remainder of this Article, we briefly highlight four major
advances and central new insights gained from our work.

(i)~We find an analytical expression for the uniform static spin susceptibility of the intrinsic
(undoped) system,
\begin{equation}\label{eq:Intrinsicq0}
\overline{\chi}^{(\text{int})}_{xx,zz}(\gamma; \qq = 0)= -\frac{\mathcal{C}^2_{x,z}(\gamma)}{16\pi |B|}
\, \frac{1}{1+4\xi_{\text{M}}\Theta(\xi_{\text{M}})} \quad ,
\end{equation}
where $\xi_{\text{M}}$ is proportional to the band gap and positive (negative) for the topologically
trivial normal (topologically nontrivial inverted) regime, $\Theta$ denotes the Heaviside step function,
$B$ is a band-structure parameter from the BHZ model introduced further below, and the constants
$\mathcal{C}_{x,z}(\gamma)$ depend both on the valence-band mixing of quantum-well basis states
and on the parameter $\gamma$ that characterises the relative coupling strength of the
conduction-band and valence-band spin degree of freedom to the physical quantity of interest.
While the intrinsic spin response is suppressed with increasing band gap in the normal regime
($\xi_M>0$), it becomes independent of gap size in the topological regime where $\xi_M \le 0$.
This unexpected behavior is a special feature of the symmetry-protected topological phase of the bulk
system associated with the existence of gapless edge states.

\begin{figure}[t]
\includegraphics[width=0.8\columnwidth]{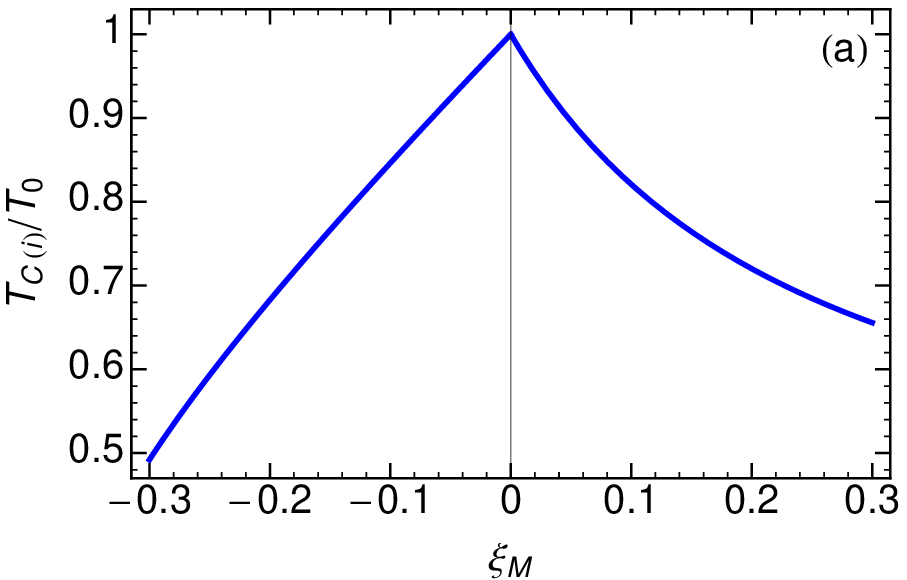}
\includegraphics[width=0.8\columnwidth]{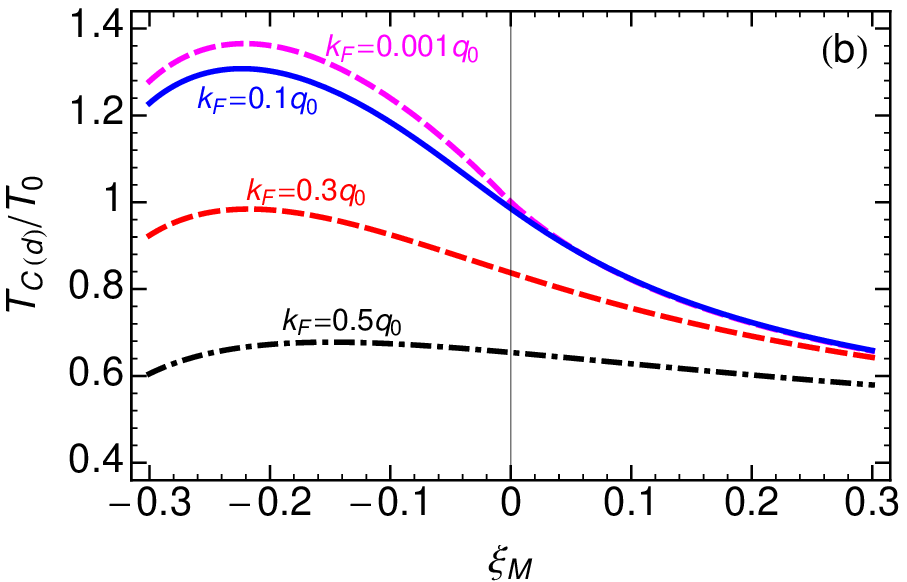}
\caption{\label{fig:CT1}
(a)~Mean-field critical temperature $T_\mathrm{C(i)}$ for virtual-carrier-mediated magnetism in
a HgTe quantum well plotted as a function of the band-gap parameter $\xi_{\text{M}}$. Note the
striking asymmetry between the topological ($\xi_M<0$) and normal ($\xi_M>0$) regimes. An
explicit expression for $T_0$ is given in the text; it contains the local exchange coupling between
magnetic-impurity orbitals and HgTe quantum-well basis states, which is roughly constant across
the transition~\cite{Dobrowolski}. (b)~Critical temperature $T_\mathrm{C(d)}$ for the electron-doped
HgTe quantum well plotted as a function of $\xi_{\text{M}}$ and for various charge-carrier densities
(corresponding to indicated values for the Fermi wave vector $k_\mathrm{F}$; where $k_\mathrm{F}
=q_0$ corresponds to a charge density $n_0\equiv q_0^2/(2\pi) = 4.25\times 10^{12}\,$cm$^{-2}$ for
a typical structure~\cite{Rothe2010NJP}). We used $\xi_{\text{D}}=-0.7$ for the BHZ-model
electron-hole-asymmetry parameter in both plots, and $\gamma=-2.22$ as relevant for the
(Hg,Cd,Mn)Te system~\cite{Dobrowolski}.}
\end{figure}

(ii)~A physical consequence of the unusual spin response in the intrinsic system is the
asymmetric variation of the critical (Curie) temperature for virtual-carrier-mediated magnetic order in a
HgTe quantum well that has been doped magnetically but not electronically. See Fig.~\ref{fig:CT1}(a).
A similarly striking asymmetry arises when charge carriers are present in the 2D conduction band;
which is illustrated in Fig.~\ref{fig:CT1}(b). Furthermore, in this situation, a rather strong, and
counter-intuitive, suppression of the Curie temperature with the density of itinerant charge carriers
is revealed. This tendency arises from the unconventional character of conduction-band states in
the topological regime. See Section~\ref{sec:dopedSR} for details.

(iii)~We extract the effective \textit{g} factors for 2D conduction electrons in both the topological
and normal regimes, which are directly accessible experimentally~\cite{Hart2015arXiv}. Their strong density
and gap-parameter dependences reflect the band mixing and transition between regimes dominated by
Schr\"odinger-type and Dirac-like dynamics of charge carriers. Full details can be found in
Section~\ref{sec:2Dgfac}. These results are essential, e.g., to enable quantitative characterization of the
transition between quantum spin-Hall and quantum anomalous Hall phases in magnetic 2D topological
insulators~\cite{Liu2008PRLb}.

(iv)~We predict the effective \textit{g} factors of the helical edge states that are present in the
topological regime of the HgTe quantum well. For the out-of-plane magnetic-field direction, the \textit{g}
factor assumes a constant value that is determined by band mixing in the quantum-well eigenstates. In
contrast, the in-plane \textit{g} factor has a strong density dependence. See Section~\ref{sec:gfactEdge}.

In the following, we discuss relevant details of our theoretical analysis and present complete results
for the spin response in both intrinsic and electron-doped systems.

\section{Quantum-well bandstructure\label{HgTeQW}}

The electronic properties of HgTe quantum wells are adequately captured by an effective four-band
(BHZ) Hamiltonian~\cite{Bernevig2006} that acts in the low-energy subspace spanned by basis
states $|E_1+\rangle, |H_1+\rangle, |E_1-\rangle, |H_1-\rangle$ and explicitly reads
\begin{eqnarray}
\mathscr{H}_0=
\begin{pmatrix} 
{\mathcal H}(\kk) & 0  \\[2mm]
0& {\mathcal H}^*(-\kk) 
\end{pmatrix}~,
\label{eq:BHZ}
\end{eqnarray}
with ${\mathcal H}(\kk)=h_\mu\sigma^\mu$, $h=(C-Dk^2,A k_x,A k_y, M-B k^2)$ and $\sigma^\mu
=(\openone,\ssig)$. The parameters $A, B, C, D, M$ are functions of the well width $d$, and their
numerical values are typically obtained by a fitting procedure~\cite{Muehlbauer2014PRL}. The
parameter $M$ opens a band gap, where (in the convention $B<0$) the system is in the topological
(normal) regime when $M<0$ ($M>0$).  The basis functions $|E_1\pm\rangle$ are a superposition
of conduction-electron and light-hole (LH) basis functions with a given spin projection. The
associated band $|L_1\pm\rangle$ has a much lower band-edge energy and can therefore be
omitted. On the other hand, the heavy-hole (HH) states $|H_1\pm\rangle$ also belong to the set of
low-energy excitations. In the following we set $C=0$, and employ a dimensionless description of
Eq.~(\ref{eq:BHZ}) that is obtained  by defining an energy scale $E_0\equiv A q_0$ and a scale for
the wave vector $q_0\equiv A/|B|~$\cite{Juergens2014PRL,Juergens2014PRB}. (In a typical HgTe
quantum-well structure~\cite{Rothe2010NJP}, $E_0=0.189$~eV and $q_0=0.517$~nm$^{-1}$.)
By making use of the axial symmetry of the BHZ Hamiltonian (\ref{eq:BHZ}) we rotate to a real basis
to obtain $\mathcal{H}'(\kk)=U^{(+)\dagger}_{\phi_\kk}\mathcal{H}(\kk)U^{(+)}_{\phi_\kk}=E_0 h'_\mu
\sigma^\mu$, with $U^{(s)}_{\phi_\kk}={\text{diag}}(\ee^{i s\phi_\kk/2},\ee^{-i s\phi_\kk/2})$ and
$h'=(-\xi_{\text{D}}\tilde k^2,\tilde k, 0, \xi_{\text{M}}+ \tilde k^2)$, where $\phi_\kk$ is the polar angle
of the 2D wave vector $\kk$. We have defined the dimensionless parameters $\xi_{\text{M}} \equiv
M/E_0$ and $\xi_{\text{D}}\equiv D/|B|$, which have typical
values~\cite{Bernevig2006,Rothe2010NJP,Muehlbauer2014PRL} $|\xi_{\text{M}}|, |\xi_{\text{D}}|
\lesssim 0.5$.
The eigenvectors in the complex and real basis are related via $a^{(s)}_{\kk\alpha}=U^{(s)}_{\phi_\kk} 
~a^{(s)}_{k\alpha}$, where $\alpha=\pm$ distinguishes the conduction and valence bands, which are
doubly degenerate in the quantum number $s=\pm$ for spin projection along the growth direction and
have the dispersions
\begin{equation}
E_{\kk\alpha}^{(s)}\equiv E_{k\alpha}^{(s)}
=E_0[-\tilde k^2 \xi_{\text{D}}+\alpha\sqrt{(\xi_{\text{M}}+\tilde k^2)^2+\tilde k^2}] \,\, .
\end{equation}

\section{Effective spin susceptibility of the 2D system: Intrinsic \& doped cases}

The spin susceptibility is used to characterize a system's response to spin-related external stimuli
within the framework of linear-response theory~\cite{vignalebook}. In the static limit, and for
quantum-well-confined electrons, it can be written as~\cite{Kernreiter2013PRL}
\begin{eqnarray}
&& \chi_{ij}(\vek{R}, z; \vek{R}', z')= \lim_{\eta\to 0^+} \bigg\{ -\frac{i}{\hbar} \int_0^\infty d t \,\,
\ee^{-\eta t} \nonumber \\ && \hspace{3cm}
\times \langle[S_i(\vek{R},z; t)\, , \, S_j(\vek{R}',z';0)] \rangle \, \bigg\} \, , \quad
\label{eq:Spinsus}
\end{eqnarray}
with $\vek{R}\equiv (x,y)$ and $z$ being coordinates in the 2D plane and perpendicular to it,
respectively, and $ S_j(\vek{R}, z )=\Psi^\dagger(\vek{R}, z)~\hat{S}_j~ \Psi(\vek{R}, z)$ denoting
the electron spin density measured in units of $\hbar$. For a homogeneous 2D electron system,
the spin susceptibility is most straightforwardly obtained in terms of the spatially Fourier-transformed
quantity $\chi_{ij}(\vek{q}; z, z')$ via
\begin{equation}
\chi_{ij}(\vek{R}, z; \vek{R}', z') = \int \frac{d^2 q}{(2\pi)^2} \,\, \ee^{i \vek{q} \cdot (\vek{R} -
\vek{R}')} \,\, \chi_{ij}(\vek{q}; z, z') \, .
\end{equation}

The dependence of $\chi_{ij}(\vek{q}; z, z')$ on the coordinates $z$ and $z'$ encodes the spatial profile
of the quantum-well bound states. Within the BHZ framework, the $z$-dependent part of electron wave
functions is contained in the four basis functions $|E_1\pm\rangle, |H_1\pm\rangle$. The latter are
spinors whose explicit form has been derived~\cite{Bernevig2006} within the six-band Kane-model
description~\cite{Winkler2003Book} for the charge-carrier dynamics that includes the bands with
$\Gamma_6$ and $\Gamma_8$ symmetry closest to the bulk-material's fundamental gap.
As is generally the case in multi-band systems, the spin response of electrons in a HgTe quantum well is
strongly influenced by both the in-plane dynamics described by the BHZ Hamiltonian and the nontrivial
spinor structure of the BHZ-model basis states. We therefore need to express the spin susceptibility within
the underlying six-band Kane model.

The coupling between some physical stimulus represented by a field $\vek{\mathcal F}$ and the
$\Gamma_6$ and $\Gamma_8$-band intrinsic angular-momentum degrees of freedom $\vek{\hat\sigma}$
and $\vek{\hat J}$ is most generally described by a term
\begin{equation}\label{eq:intSpinHam}
{\mathscr H}_{\vek{\mathcal F}} = \sum_i {\mathcal F}_i \left( b_{\Gamma_6}\,\frac{\hat\sigma_i}{2}
\oplus 0_{4\times 4} + b_{\Gamma_8}\, 0_{2\times 2}\oplus\hat J_i \right)
\end{equation}
in the Kane-model Hamiltonian. See, e.g., Table~C.5 in Ref.~\onlinecite{Winkler2003Book}.
Within this approach, the coefficients $b_{\Gamma_j}$ are intra-band coupling constants  with appropriately
renormalized values to take account of all field-induced band-coupling effects in the bulk material.
To be able to discuss a wide range of spin-related phenomena, we define an effective (pseudo-)spin
operator
\begin{equation}\label{eq:effKaneSpin}
\hat S_i (\gamma) = \frac{\hat\sigma_i}{2}\oplus (\gamma \hat J_i) \quad ,
\end{equation}
such that ${\mathscr H}_{\vek{\mathcal F}} \equiv b_{\Gamma_6} \sum_i {\mathcal F}_i \, \hat S_i
(b_{\Gamma_8}/b_{\Gamma_6})$. The actual value of the parameter $\gamma$ depends on the
physical quantity or situation of interest~\footnote{For example, when discussing Pauli
paramagnetism, the response function for $\hat S_i(\gamma)$ with $\gamma = -2\kappa/g_\ast$
is relevant, where $g_\ast$ and $\kappa$ are the respective $g$ factors for the $\Gamma_6$ and
$\Gamma_8$ bands in the Kane-model band-structure description~\cite{Winkler2003Book}.
In contrast, the intrinsic real-spin polarisation is associated with $\hat S_i(1/3)$. See, e.g., Eq.~(6.65)
in Ref.~\onlinecite{Winkler2003Book}. To discuss carrier-mediated magnetism, the susceptibility with
$\gamma = \beta/\alpha$ in the notation of Ref.~\onlinecite{Dobrowolski} needs to be considered.
Finally, $\hat S_i(1)$ is the proper angular-momentum operator that generates spatial rotations.}.
The system's response is then fully captured by the effective spin-susceptibility tensor
\begin{subequations}
\begin{eqnarray}\label{eq:WavevecSpinSus}
&& \chi_{ij}(\gamma; \vek{q}; z, z') = \sum_{\alpha,\beta,s,s'} \int\frac{d^2 k}{(2\pi)^2} \,\,
\mathscr{W}_{ij(\kk,\kk+\qq,\alpha,\beta)}^{(s,s')}(\gamma; z,z') \nonumber \\ 
&&\hspace{3cm} \times  \frac{n_{\text F}(E^{(s)}_{\kk\alpha})-n_{\text F}
(E^{(s')}_{\kk+\qq\beta})}{E^{(s)}_{\kk\alpha} -E^{(s')}_{\kk+\qq\beta}+i\hbar\eta}~,
\end{eqnarray}
where $n_{\text{F}}$ denotes the Fermi function, and
\begin{eqnarray}\label{eq:SpinOverlap}
&&\mathscr{W}_{ij(\kk,\kk+\qq,\alpha,\beta)}^{(s,s')}(\gamma; z,z')= \left[
\psi^{(s)}_{\kk\alpha}(z)\right]^\dagger \cdot \Big[\hat{S}_i (\gamma) \,
\psi^{(s')}_{\kk+\qq\beta}(z)\Big] \nonumber \\
&& \hspace{2cm} \times \Big[\psi^{(s')}_{\kk+\qq\beta}(z')\Big]^\dagger \cdot \Big[
\hat{S}_j(\gamma) \, \psi^{(s)}_{\kk\alpha}(z')\Big]~,
\end{eqnarray}
\end{subequations}
are matrix elements of the effective spin operator given in Eq.~(\ref{eq:effKaneSpin}). In the spirit of
subband $\kk\cdot\vek{p}$ theory, the six-dimensional spinor wave functions $\psi^{(s)}_{\kk\alpha}(z)$
can be expressed in terms of the BHZ-model basis state spinors $\psi^{(s)}_{0i}(z)$ as
\begin{equation}\label{eq:kdepspinors}
\psi^{(s)}_{\kk\alpha}(z)=
\sum^2_{i=1}  
\left(U^{(s)}_{\phi_\kk}\right)_{ii}a_{k\alpha, i}^{(s)}~\psi_{0i}^{(s)}(z)~,
\end{equation} 
where the coefficients $a_{k\alpha, i}^{(s)}$ are the components of the corresponding eigenvectors of
the BHZ Hamiltonian (see Sec.~\ref{HgTeQW}).
The explicit form of the basis states was derived in Ref.~\onlinecite{Bernevig2006} by solving a
confined-particle problem in the HgTe/CdTe hybrid system. For instance, $\psi^{(+)}_{01}(z)^T=
(f_1(z),0,0,f_4(z),0,0)$, with the detailed form of the envelope function components $f_i(z)$
provided in the supplemental information of Ref.~\onlinecite{Bernevig2006}.

In the following, we consider the growth-direction-averaged spin susceptibility of charge carriers in
the HgTe quantum well, given by $\overline{\chi}_{ij}(\gamma;\vek{q}) = \int d z \int dz' ~ \chi_{ij}
(\gamma; \vek{q}; z, z')$~\cite{Kernreiter2013PRL}, with $\chi_{ij}(\gamma; \vek{q}; z, z')$ calculated
using the Kane-model-based BHZ approach as described above. Note that, by replacing the spin
matrices in Eq.~(\ref{eq:SpinOverlap}) by the unit matrix and averaging over $z,z'$, we obtain the
charge-response function studied in Refs.~\onlinecite{Juergens2014PRL,Juergens2014PRB}. From
the axial symmetry of our Kane-model description, and the fact that the eigenstates have definite spin
projection in the growth direction, it follows  that the in-plane spin susceptibilities are the same, i.e.,
$\overline{\chi}_{xx}(\gamma;q)=\overline{\chi}_{yy}(\gamma;q)$, and only depend on the magnitude
$q\equiv |\qq|$. This is an important difference to other semiconducting systems where HH-LH mixing
occurs~\cite{Kernreiter2013PRL}. In the present situation, HH-LH mixing arises only from terms linear
in $\kk$ (giving rise to Dirac-like excitations) which is a consequence of the envelope function
components~\cite{Bernevig2006} behaving differently (even or odd) under the parity transformation
$z\leftrightarrow -z$.

\subsection{Spin response of the intrinsic system}

\begin{figure}[b]
\includegraphics[width=0.8\columnwidth]{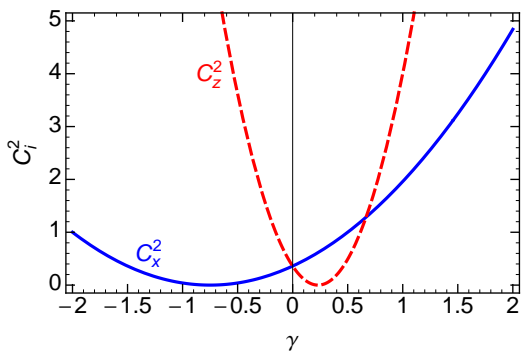}
\caption{\label{fig:Ci}
Prefactors $C_x^2$ and $C_y^2$ determining the magnitude of intrinsic in-plane and out-of-plane 
spin response, respectively [see Eq.~(\ref{eq:Intrinsicq0})], plotted as a function of the parameter
$\gamma$ from the definition of the effective Kane-model spin operator (\ref{eq:effKaneSpin}).}
\end{figure}

In the insulating limit, the conduction band is empty and the Fermi level lies in the band gap, i.e.,
$|\mu|<E_0|\xi_{\text{M}}|$. In this situation, the spin susceptibility originates from virtual interband
transitions across the band gap. In the following we will focus on the zero temperature limit. The
analytical result for the intrinsic spin susceptibility obtained in the limit of zero momentum transfer
$q\to 0$ is given in
Eq.~(\ref{eq:Intrinsicq0}), where
\begin{subequations}\label{eq:coeffCi}
\begin{eqnarray}
{\mathcal C}_x(\gamma) &=& 1 + (2 \gamma - 1)\, \mathcal{C}_{\text{LH}} \quad , \\
{\mathcal C}_z(\gamma) &=&  1-3\gamma  + (\gamma - 1)\, \mathcal{C}_{\text{LH}} \quad ,
\end{eqnarray}
\end{subequations}
and $\mathcal{C}_{\text{LH}}\equiv \int dz |f_4(z)|^2$ ($\approx 0.4$) is the amount of LH admixture
in the basis states $|E_1\pm\rangle$. (Details about the calculation of the intrinsic spin susceptibility
are given in Appendix~\ref{sec:IntrinsicNum}. Results for $q\neq0$ can be calculated numerically. For
completeness, some of these are also shown in Appendix~\ref{sec:IntrinsicNum}.) Equation~(\ref{eq:Intrinsicq0})
exhibits anomalous behavior in the inverted region ($\xi_{\text{M}}<0$) in that the uniform spin response is
independent of the gap size and pinned to the value for the gapless case, even though it arises from virtual
inter-band transitions. In the normal region ($\xi_{\text{M}}>0$), the expected decrease of the response
functions with increasing band gap is found. The spin susceptibility  (\ref{eq:Intrinsicq0}) is strongly
anisotropic, generally exhibiting a dominant out-of-plane response except for a very small range of the
parameter $\gamma$. See Fig.~\ref{fig:Ci}.

A non-vanishing $\overline{\chi}^{(\text{int})}_{jj}(\gamma;0)$ seems to imply
the counterintuitive phenomenon that an applied magnetic field could generate a magnetisation of
the HgTe quantum-well system in the intrinsic limit where no charge carriers are present. Our more
detailed analysis shows, however, that this is not the case. Direct calculation of the derivative of
the free energy with respect to magnetic-field strength in a model based on the BHZ Hamiltonian
(\ref{eq:BHZ}) augmented by a Zeeman term reveals that the total magnetisation is strictly zero
\footnote{For instance, for a perpendicular magnetic field ${\vek{\mathcal B}}={\mathcal
B}_z\, {\mathbf{\hat z}}$, the grand canonical potential is 
given by $J(T,{\mathcal B}_z,\mu)=-\frac{L^2}{\beta} \sum_\nu\int\frac{d^2k}{(2\pi)^2}\ln\left(1+
\ee^{-\beta[\varepsilon_\nu(k,{\mathcal B}_z)-\mu]}\right)$, where $\beta^{-1}=k_{\text{B}}T$ and
$\varepsilon_\nu(k,{\mathcal B}_z)$ are the dispersions of the two spin-split bands ($\nu=1,2$)
obtained from the effective Hamiltonian. In the limit $T=0$ (and $\mu=0$), we obtain
$J({\mathcal B}_z)=\frac{L^2}{2\pi}\sum_\nu\int_0^\Lambda dk~k~\varepsilon_\nu(k,{\mathcal B}_z)$.
The net-magnetisation of the bands is given by $\mathcal{M}_z=\frac{1}{L^2}\frac{\partial J({\mathcal 
B}_z)}{\partial {\mathcal B}_z}$. Numerically we find that  $\frac{\partial J({\mathcal B}_z)}{\partial
{\mathcal B}_z} \propto \sum_\nu c_\nu(\Lambda)=0$, since the numerical coefficients of the two
spin-split bands are related as $c_1(\Lambda)=-c_2(\Lambda)$.}. This is due to the fact that
intrinsic-spin and orbital contributions to the magnetization cancel, as expected in a spin-orbit-coupled 
system~\footnote{In other words, while the total \emph{magnetic\/} susceptibility of the HgTe
quantum-well system is trivial (i.e., vanishes) in the intrinsic limit, the \emph{effective-spin\/}
susceptibility is finite and exhibits interesting properties that affect observable physical phenomena,
e.g., impurity-spin exchange and Zeeman splitting; as discussed in the following.}. This conclusion is
further underpinned by the observation that $\overline{\chi}^{(\text{int})}_{jj}(\gamma;0)$ vanishes in
the limit of zero  HH-LH mixing~\footnote{The switching off of HH-LH mixing can be simulated by
replacing  $A\to\lambda A$ in Eq.~(\ref{eq:BHZ}) and letting $\lambda\to 0$ after integration}.

\begin{figure}[t]
\includegraphics[height=4cm]{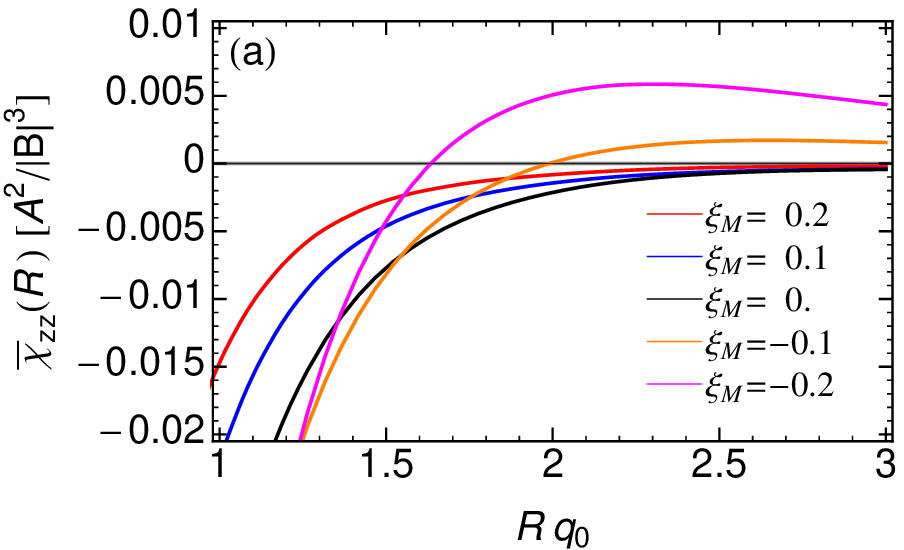}
\includegraphics[height=4cm]{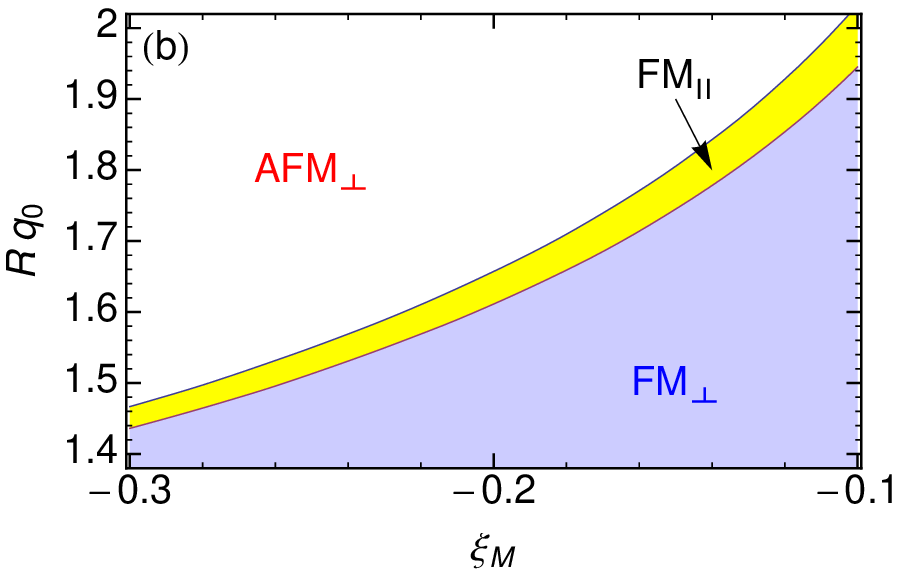}
\caption{\label{fig:IntchizzR}
(a) Intrinsic real-space spin susceptibility $\overline\chi_{zz}(R)$ plotted as a function of $R\, q_0$
for several values of the gap parameter $\xi_{\text{M}}$, with fixed $\xi_{\text{D}}=-0.7$ and
$\gamma=-2.22$. (b) Parameter domains of preferential alignments between localized Mn spins
having a distance $R$, mediated by the intrinsic spin response of an HgTe quantum-well system.
The blue (white) domain favours a ferromagnetic (anti-ferromagnetic) out-of-plane
spin alignment. In the yellow region the two localized spins tend to align ferromagnetically in-plane.}
\end{figure}

As it is possible to engineer and study effective exchange interactions between impurity
atoms~\cite{Zhou2010NP,Khajetoorians2012NP}, it is tempting to consider such interactions between
two localized spins in a HgTe quantum well. Quite generally, the RKKY Hamiltonian is given
by~\cite{yosidabook}
\begin{equation}\label{eq:RKKYHam}
\mathscr{H}^{\text{eff}}_{\vek{r},\vek{r}'}=G^2\sum_{i,j} ~I^{i}_{\vek{r}}~I^{j}_{\vek{r}'}~\chi_{ij}(\vek{r},
\vek{r}').
\end{equation}
Here $\chi_{ij}(\vek{r},\vek{r}')$, the Fourier transform of (\ref{eq:WavevecSpinSus}), is the spin
susceptibility in real space, $I^{i}_{\vek{r}}$ denotes the $i$th Cartesian component of an impurity
spin located at position $\vek{r}=(\vek{R},z)$, and $G$ is the local exchange-coupling constant 
between the spin degree of freedom carried by band electrons and localized (e.g., impurity) spins.
(The difference in exchange-coupling strengths for conduction-band and valence-band states is
accounted for by the appropriate value of $\gamma$.) In Fig.~\ref{fig:IntchizzR}(a), we plot the
growth-direction-averaged effective out-of-plane real-space spin susceptibility $\overline{\chi}_{zz}
(R)$ as a function of the distance $R$ for various values of $\xi_{\text{M}}$ (and $\xi_{\text{D}}=
-0.7$ corresponding to realistic situations~\cite{Muehlbauer2014PRL}).

Figure~\ref{fig:IntchizzR} shows that the effective exchange interaction mediated by the intrinsic
spin response of the HgTe quantum well for localized spins is of ferromagnetic (FM) type if $R q_0
\ll 1$, and that there is a cross-over to antiferromagnetic (AFM) coupling for $R q_0\gtrsim 1$ that
sensitively depends on $\xi_{\text{M}}$. In particular, we find that this cross-over is shifted significantly
to smaller impurity-spin separations in the case of the inverted regime as compared to the normal
regime. Also, the magnitude of exchange interaction in the ferromagnetic regime for fixed distance
$R$ markedly increases when enlarging the gap in the inverted system. In contrast, the exchange
interactions in the normal regime have a very small magnitude.
To further illustrate the parametric dependencies of the carrier-mediated exchange interaction,
Fig.~\ref{fig:IntchizzR}(b) shows the various regions of preferential spin alignments as a function
of $\xi_{\text{M}} < 0$. Interestingly, inbetween the two out-of-plane localized-spin alignments
(FM${}_\perp$ and AFM${}_\perp$), we also find sizeable regions for the distance where it is
energetically favorable for impurity spins to align ferromagnetically in-plane (FM${}_\parallel$). 
Moreover, for larger distances, we find the usual Bloembergen-Rowland~\cite{Bloembergen1955PRB}
behavior for the intrinsic spin susceptibilities which is characterized by an exponential decay factor
that depends on the band gap, approximately given by $\sim \exp{(-R/R_0)}/R^3$ where $R_0=
(|\xi_{\text{M}}|q_0)^{-1}$ is the Compton wave length of the band electrons. More details are given
in Appendix~\ref{sec:BloembergenRowland}. In the limit of zero gap ($\xi_{\text{M}}\to 0$), the
$R^{-3}$ decay found previously in various Dirac systems~\cite{Stauber2003PRB,Liu2009PRL,Zhu2011PRL}
is reproduced.

Having considered the case of two localized impurity spins with exchange interactions mediated by
virtual excitations, we now focus on a HgTe quantum well that is doped with a large number of
homogeneously distributed magnetic impurities (eg, Mn ions), forming effectively a
Hg${}_{1-x}$Mn${}_x$Te alloy~\cite{Bastard1979PRB}. The average distance between the spins is
$R_{\text{NN}}\approx (3 a^3_0/16\pi x)^{1/3}$~\cite{Litvinov2001PRL}, with $a_0$ the HgTe lattice
constant and $x$ the concentration of magnetic ions. We calculate the Curie temperature in the mean
field limit, assuming $R_{\text{NN}}/R_0\ll1$ which is justified when $x\gg 2\times 10^{-3}
|\xi_{\text{M}}|^3$, where we have used $q_0\lesssim 0.5$~nm${}^{-1}$. Since $|\xi_{\text{M}}|\lesssim
0.3$~\cite{Muehlbauer2014PRL}, this condition is practically always fulfilled. The Curie temperature for
Ising-type ferromagnetic order with magnetization perpendicular to the quantum well growth direction is
given by
\begin{eqnarray}\label{eq:TCMF}
T_\mathrm{C(i)} = T_0~\frac{d_{\text{c}}/d(\xi_{\text{M}})}{1+4\xi_{\text{M}}
\Theta(\xi_{\text{M}})}~,
\end{eqnarray}
where $T_0=\frac{I(I+1){\mathcal C}^2_z(\gamma)}{48\pi}\frac{G^2}{k_{\text{B}}|B|}
\frac{n_I}{d_{\text{c}}}$, $I$ denotes the impurity-spin magnitude, $n_I$ is the 3D density of magnetic
impurities, and $d_{\text{c}}\approx 6.3$~nm is the critical well width. For obtaining Eq.~(\ref{eq:TCMF})
we have used the approximation $\int_{-d/2}^{d/2} dz \int_{-d/2}^{d/2} dz' \chi^{(\text{int})}_{jj}(\gamma;
0,z,z')\approx \overline{\chi}^{(\text{int})}_{jj}(\gamma; 0)$. In Fig.~\ref{fig:CT1}(a), we show the Curie
temperature as a function of $\xi_{\text{M}}$, where we set $A=0.375$~eV~nm and
$B=-1.120$~eV~nm${}^2$~\cite{Muehlbauer2014PRL} as their dependence on $d$ is much weaker
compared to $M$. We see that the behavior of the Curie temperature in its dependence on
$\xi_{\text{M}}$, or equivalently $d$, provides a clear means to distinguish between topological and
normal regions.
 
Knowing the spin susceptibility as a function of the wave vector allows us to go beyond the mean
field limit~\cite{Simon2007PRL,Simon2008PRB} and discuss the stability of the mean-field ground 
state with respect to thermally excited spin waves (magnons). In our present case of interest, the
magnon dispersion for $q/q_0\ll1$ is given by $\omega_q=\omega_0+c_2q^2$, with a coefficient
$c_2=\left. \frac{1}{2}\frac{\partial^2 \overline{\chi}^{(\text{int})}_{zz}(\gamma;q)}{\partial q^2}
\right|_{q=0}$. Using Figure~\ref{fig:Spinsusq} in Appendix~\ref{sec:IntrinsicNum}, which shows
results for the relevant set of parameters, we find $c_2\lesssim 0$. Thus the
criterion~\cite{yosidabook,Simon2008PRB,Kernreiter2013PRL} $c_2>0$ needed to guarantee
stability of the mean-field ground state is generally violated. It may be possible that electron-electron
interactions and/or spin-orbit coupling help to stabilize ferromagnetic order, as has been shown to be
the case for an ordinary 2D electron gas~\cite{Simon2007PRL,Simon2008PRB,Zak2012PRB}, but
this question is beyond the scope of our present work.

\subsection{Spin response of the electron-doped system\label{sec:dopedSR}}

We consider now situations where the conduction
band is occupied ($\mu>E_0|\xi_{\text{M}}|$). In the $q\to 0$ limit, an analytical expression for
the extrinsic contribution to the spin susceptibility (arising from filled conduction-band states) is
found as
\begin{widetext}
\begin{subequations}\label{eq:Chijjq0}
\begin{eqnarray}
\overline{\chi}_{xx}^{(\text{dop})}(\gamma;0)&=&\frac{\mathcal{C}^2_x(\gamma)}{16\pi|B|}\Biggl[\frac{2\pi
\tilde N(0)}{f^2}\Big(\tilde k_{\text{F}}^2 - 2 f (\tilde k_{\text{F}}^2 +\xi_{\text{M}}) -\, 2 f^2\Big) +
\frac{\tilde k_{\text{F}}^2 (1+ 2 \xi_{\text{M}}) + 2 \xi_{\text{M}}^2 - 2 f |\xi_{\text{M}}|)}{f(1 + 4 
\xi_{\text{M}})}\Biggr]~,\\[2mm]
\overline{\chi}_{zz}^{(\text{dop})}(\gamma;0)&=&\frac{\mathcal{C}^2_z(\gamma)}{16\pi|B|}\Biggl[\frac{2\pi
\tilde N(0)}{f^2}\left(\frac{6 \gamma \mathcal{D}_z(\gamma)}{\mathcal{C}^2_z(\gamma)}\,
\tilde k_{\text{F}}^2 - \frac{4 f^2 \{\mathcal{D}^2_z(\gamma)(\tilde k_{\text{F}}^2 +\xi_{\text{M}}+f)^4
+ 9\gamma^2\tilde k_{\text{F}}^4\}}{\mathcal{C}^2_z(\gamma)[\tilde k_{\text{F}}^2+(\tilde k_{\text{F}}^2
+\xi_{\text{M}}+f)^2]^2}\right) +\frac{\tilde k_{\text{F}}^2 (1+ 2 \xi_{\text{M}}) + 2 \xi_{\text{M}}^2 -
2 f |\xi_{\text{M}}|)}{f(1 + 4  \xi_{\text{M}})}\Biggr]~, \nonumber \\
\end{eqnarray}
\end{subequations}
\end{widetext}
where the first (second) term between the square brackets in the expression for
$\overline{\chi}_{jj}^{(\text{dop})}(\gamma;0)$ is an intraband (interband) contribution. In 
Eqs.~(\ref{eq:Chijjq0}), we used the abbreviations $\mathcal{D}_z(\gamma)\equiv
(1-\gamma)\,\mathcal{C}_{\text{LH}}-1$ and $f\equiv \sqrt{(\tilde k_{\text{F}}^2+\xi_{\text{M}})^2
+\tilde k_{\text{F}}^2}$. $\tilde N(0)$ is related to the density of states at the Fermi energy as
$-\overline{\chi}_{0}(0)=N(0) = (2/|B|)~ \tilde N(0)$ (where $\overline{\chi}_{0}(q)$ is the static
charge-response function), and reads $\tilde N(0)=\frac{1}{2\pi}\left|\frac{1+2\tilde k_{\text{F}}^2
+2\xi_{\text{M}}}{\sqrt{(\tilde k_{\text{F}}^2+\xi_{\text{M}})^2+\tilde k_{\text{F}}^2}}-2\xi_{\text{D}}
\right|^{-1}$. For the general case with $q>0$, $\overline{\chi}_{jj}(\gamma;q)$ can be calculated
numerically and its line shape is presented in Appendix~\ref{sec:DopedNum} for various examples.
Due to the sharpness of the Fermi surface, we find for the corresponding real-space spin susceptibilities
the expected $R^{-2}$ oscillatory decay expected from a 2D Fermi liquid~\cite{vignalebook}.
 
Next we calculate the Curie temperature using the mean field approach under the premise
$k_{\text{F}}R_{\text{NN}}\ll1$, which leads to the condition $x\gg 2\times 10^{-3}~\tilde k_{\text{F}}$.
This condition is generally fulfilled since $k_{\text {F}}\lesssim 0.1$~nm${}^{-1}$ is the reliable range  
of the effective model~\cite{Muehlbauer2014PRL}. The Curie temperature is given by
\begin{eqnarray}
T_\mathrm{C(d)} = T_0~\frac{d_{\text{c}}}{d(\xi_{\text{M}})}~ \frac{16\pi|B|}{C^2_z(\gamma)}
~|\overline{\chi}_{zz}(\gamma;0)|~,
\end{eqnarray}
where $\overline{\chi}_{zz}(\gamma;0)=\overline{\chi}^{(\text{int})}_{zz}(\gamma;0)+
\overline{\chi}^{(\text{dop})}_{zz}(\gamma;0)$. Note that also in the doped case Ising type magnetism 
prevails. In Fig.~\ref{fig:CT1}(b), we show the Curie temperature as a function of $\xi_{\text{M}}$ for
various values of the Fermi wave vector. For low doping, we see a strong dependence of the Curie
temperature on $\xi_{\text{M}}$ in the topological regime, which becomes very weak in the normal
regime. For large doping, on the other hand, the Curie temperature is not very sensitive to
$\xi_{\text{M}}$ in its whole range. Furthermore, the Curie temperature is generally suppressed with
increased doping, but this trend is much stronger in the topological regime.
For the doped case, we find the magnon dispersion $\omega_q=\omega_0+\bar{c}_2q^2$ where
$\bar{c}_2\equiv\frac{1}{2}\left.\frac{\partial^2 \overline{\chi}_{zz}(\gamma;q)}{\partial q^2}\right|_{q=0}
>0$ always, in contrast to the intrinsic case. See Fig.~\ref{fig:SpinSusdop} in
Appendix~\ref{sec:DopedNum}. Thus a necessary condition for the stability of the ferromagnetic
groundstate is fulfilled~\cite{yosidabook,Simon2008PRB,Kernreiter2013PRL}.

\section{Confinement dependence of the 2D-electron effective $\mathbf g$ factor\label{sec:2Dgfac}}

The paramagnetic response of charge carriers in quantum-confined structures is usually
interpreted in terms of the Pauli spin susceptibility and quantified by an effective single-particle
$g$ factor~\cite{Tutuc2002PRL,Zhu2003PRL,Sashkin2006PRL}. However, any actually
measured spin-related quantities correspond almost always to averaged collective responses
of the, e.g., quasi-2D, electron system that can be crucially affected by nontrivial spin-related
phenomena~\cite{Kernreiter2013PRL}. Here we consider the paramagnetic response of
conduction-band electrons in HgTe quantum wells and show how their paramagnetic response
is changed as a function of the band-gap parameter that drives the transition between the
topological and normal regimes.

To define an effective $g$-factor for our system of a HgTe quantum well, we introduce the bulk-material
Zeeman term $\mathscr{H}_{\vek{\mathcal B}}=g_\ast\, \mu_{\text{B}}\sum_j {\mathcal B}_j\, \hat S_j
(-2\kappa/g_\ast)$, where $\mu_{\text{B}}$ is the Bohr magneton and $g_\ast$ ($\kappa$) the
$\Gamma_6$-band ($\Gamma_8$-band) $g$ factor~\cite{Winkler2003Book}. Linear-response theory
enables us to determine the paramagnetic response to the magnetic field, which is given by
$\overline{\chi}_{\text{P},j} = (g_\ast \mu_{\text{B}})^2 \, \overline{\chi}^{(\text{dop})}_{jj}(-2\kappa/
g_\ast; q)|_{q=0}$, where $\overline{\chi}_{jj}^{(\text{dop})}(-2\kappa/g_\ast;q)$ are the spin
susceptibilities of the electron doped system for the in-plane and out-of-plane response involving
$\hat S_j(-2\kappa/g_\ast)$, see Eq.~(\ref{eq:Chijjq0}) for $\gamma=-2\kappa/g_\ast$. We compare
this with the Pauli susceptibility given by $\overline{\chi}_{\text{P},j} =(g_j \mu_{\text{B}}
)^2~\overline{\chi}_{0}(0)/4$, with $\overline{\chi}_{0}(0)$ being (up to a minus sign) the density of
states which is the zero-$q$ limit of the Lindhard function and $g_j$ is the Land\'e $g$-factor for the
two directions. Thus, we can extract collective $g$ factors for the charge carriers as
\begin{equation}\label{eq:gfactor}
g_j = g_\ast \,\, \sqrt{\left. 4\,\, \frac{\overline{\chi}^{(\text{dop})}_{jj}(-2\kappa/g_\ast;
q)}{\overline{\chi}_{0}(q)} \right|_{q=0}} \quad .
\end{equation}
Our approach to determine $g$ factors via the spin susceptibility complements previous
work~\cite{Koenig2008JPSJ} where an effective Zeeman term was derived for the BHZ Hamiltonian.

\begin{figure}[b]
\includegraphics[width=0.8\columnwidth]{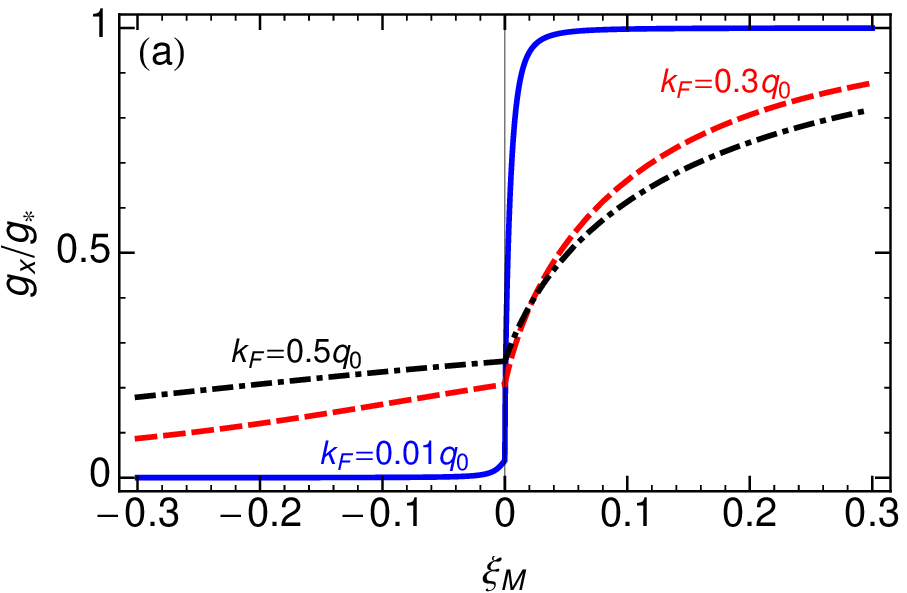}
\includegraphics[width=0.8\columnwidth]{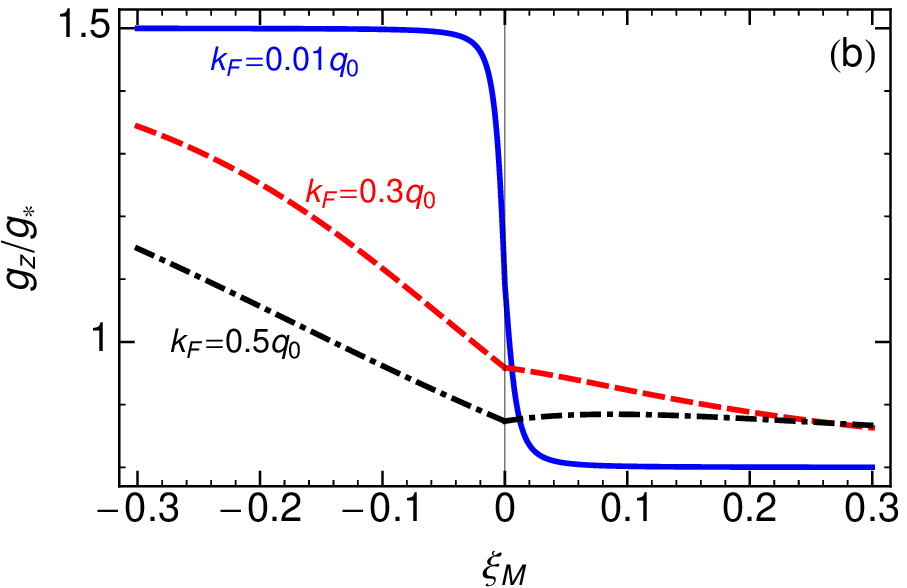}
\caption{\label{fig:g-factor}
Effective $g$-factors associated with (a) the in-plane and (b) the out-of-plane response of an
electron-doped HgTe quantum well, plotted as a function of the gap parameter $\xi_{\text{M}}$
for various levels of doping and with $\xi_{\text{D}}=-0.7$, $2\kappa/g_\ast = -0.5$. (Note that
$k_\mathrm{F}=q_0$ corresponds to a charge density $n_0\equiv q_0^2/(2\pi)= 4.25\times
10^{12} \,$cm$^{-2}$ for a typical structure~\cite{Rothe2010NJP}.)}
\end{figure}

For a narrow-gap semiconductor, the values for $g_\ast$ and $\kappa$ are dominated by
band-coupling contributions. Using generic expressions resulting from Kane-model
descriptions~\cite{Winkler2003Book}, we find
\begin{equation}\label{eq:KaneGfact}
\frac{2\kappa}{g_\ast} = -\frac{\xi_M + \xi_\Delta}{2\xi_\Delta} \quad ,
\end{equation}
with $\xi_\Delta = \Delta_0\, |B|/A^2$ in terms of the spin-orbit-splitting gap $\Delta_0$ between the
$\Gamma_8$ and $\Gamma_7$ band edges in the $8\times 8$ Kane model. As $\xi_\Delta \gg |
\xi_M|$ for our situations of interest, we can set $2\kappa/g_\ast\to -0.5$ in the following. Hence, in
Fig.~\ref{fig:g-factor}, we show the effective $g$-factors for in-plane and out-of-plane responses as a
function of $\xi_{\text{M}}$ for various levels of electron doping obtained for $2\kappa/g_\ast = -0.5$.

For very low densities, we see a vanishing in-plane response and a maximal $g_z = 6 \kappa$
HH-like out-of-plane response in the topological region $\xi_{\text{M}}<0$. This behavior arises
because the conduction-band character is dominated by the HH basis states, which experience 
a ``frozen'' spin orientation perpendicular to the quantum well due to the confinement-induced HH-LH 
energy splitting~\cite{Winkler2008SST}.  This can be easily verified from (\ref{eq:Chijjq0}) by taking
the limit $k_{\text{F}}\to 0$. Departures from these results occur for larger doping levels (larger
$k_{\text{F}}$) due to increased HH-LH mixing. In the normal region ($\xi_{\text{M}}>0$), we
encounter the situation that the conduction band is dominantly composed of the electron basis states,
rendering the \textit{g} factor to be sizable at any doping. The fact that the in-plane spin response is
notably larger than the out-of-plane response for $\xi_{\text{M}}>0$ and low doping is due to the LH
admixture in the conduction-band states. As the carrier density increases, the concomitantly increased
HH-LH mixing results in significant modifications. More detailed exploration of parametric
dependences exhibited by the collective \textit{g} factors obtained here will be useful to augment 
previous perturbative estimates~\cite{Koenig2008JPSJ} and aid the interpretation of recent
measurements~\cite{Hart2015arXiv}.

\section{Spin response of quasi-1D helical edge states in the topological regime\label{sec:gfactEdge}}

We have also investigated the edge-state contributions to the spin susceptibility in the topological
region, finding them to be negligible compared to the bulk contributions in almost all situations.
The only exception occurs for the in-plane response function $\overline{\chi}_{xx}(\gamma;q)$ in the 
purely intrinsic situation where the Fermi level lies in the minigap of the edge-state dispersions that
opens up in a finite-size sample~\cite{Zhou2008PRL,Qi2011RMP}. More details and full results
pertaining to edge states are given in Appendices \ref{sec:compbulkedge} and \ref{sec:spinsusedge}.

\begin{figure}[b]
\includegraphics[width=0.8\columnwidth]{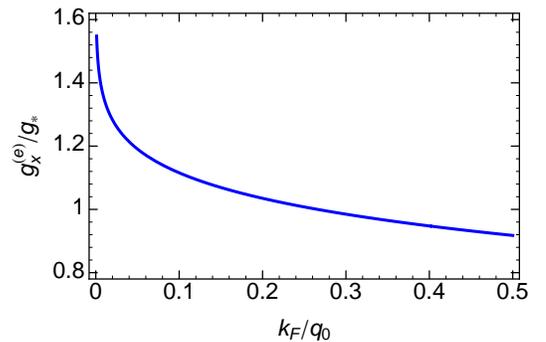}
\caption{\label{fig:g-factoredge}
Effective $g$-factor characterising the response of helical edge states to an in-plane magnetic
field.}
\end{figure}

Upto very small finite-size corrections, the paramagnetic response of helical edge states is captured
by using the effective $g$-factors $g^{(\text{e})}_{x,z}$ given by
\begin{subequations}\label{eq:gfactoredge}
\begin{eqnarray}
\frac{g^{(\text{e})}_x}{g_\ast} &=& \sqrt{\frac{\mathcal{C}^2_x(-2\kappa/g_\ast)}{4}\,\, \ln\left(
\frac{\Lambda}{k_{\text{F}}}\right)} \,\,\stackrel{\frac{2\kappa}{g_\ast} = -\frac{1}{2}}{\longrightarrow}
\,\, \frac{\sqrt{\ln\left(\frac{\Lambda}{k_{\text{F}}}\right)}}{2}~, \nonumber \\ \\
\frac{g^{(\text{e})}_z}{g_\ast} &=& -\frac{4\kappa}{g_\ast} + \frac{1 - {\mathcal C}_\mathrm{LH}}{2}
\left( 1 + \frac{2\kappa}{g_\ast}\right) \,\,\stackrel{\frac{2\kappa}{g_\ast} = -\frac{1}{2}}{\longrightarrow}
\,\,  1 + \frac{1 - {\mathcal C}_\mathrm{LH}}{4} \, . \nonumber \\
\end{eqnarray}
\end{subequations}
The results shown in Eqs.~(\ref{eq:gfactoredge}) were obtained using Eq.~(\ref{eq:gfactor}) together
with Eqs.~(\ref{eq:susedgezz})-(\ref{eq:susedgexx}). The prediction of the edge-state \textit{g} factor is
a major result of our work. In Fig.~\ref{fig:g-factoredge}, we plot the in-plane $g$-factor as a function of
$k_{\text{F}}$ where we used the natural cut-off scale $\Lambda=\pi/a_0$. Interestingly, the in-plane
$g$-factor of edge states decreases monotonically with increasing doping level. Such a behavior is in
stark contrast to the bulk case shown in Fig.~\ref{fig:g-factor}(a) and to other 2D
systems~\cite{Kernreiter2013PRL,Hatami2014PRB} where the in-plane response increases with
increased doping. This behavior reflects the fact that the spin-quantisation axis of the edge states is
perpendicular to the 2D plane, i.e., parallel to the $z$ direction, and their helical nature.

\section{Conclusions}

We present a detailed theoretical study of the spin response in HgTe quantum wells where
virtual-carrier-related processes are particularly relevant and exactly tractable within the
effective BHZ-model description. Anomalous properties of the spin susceptibility and carrier-mediated
magnetism are found in the inverted regime, extending our current understanding of spin-related
properties in topological materials.

Most strikingly, the uniform static spin susceptibility in the intrinsic limit is constant, independent
of the band gap, in the topological regime [see Eq.~(\ref{eq:Intrinsicq0})]. This results in a distinct
asymmetry between normal and inverted systems illustrated, e.g., by the dependence of the
critical temperature for virtual-carrier-mediated magnetic order in a system that is doped with
magnetic ions. Important differences between the two regimes are also exhibited in the situation
where the conduction band becomes filled (see Fig.~\ref{fig:CT1}). The stability of mean-field
ferromagnetic ground states with respect to thermal excitations of magnons has been analysed.
We find that the magnetic order is stable in the situation with finite doping.

In the topological regime, quasi-1D helical edge states exist. We have investigated their
spin-response properties, finding that their contribution to the spin susceptibility is thermodynamically
suppressed compared to that arising from 2D quantum-well states, \emph{except\/} in the very
special -- and probably physically hard-to-realise -- situation when the chemical potential is in the
mini-gap opened by the hybridisation of states from opposite edges in a finite sample.

We have used our results obtained for the spin susceptibility to define effective collective $g$
factors for states from the quasi-2D quantum-well subbands and also for the quasi-1D helical
edge states. The different character of quasi-2D-subband states in the normal and topological
(inverted) regimes is reflected in the values for the effective $g$ factor. Their strong dependence
on charge-carrier density reveals the importance of interband mixing. The behavior of the $g$
factors found for the edge states reflects their helical nature and spin-quantization property.

Our results are directly relevant for current and potential future experimental investigations of the
spin-related properties of 2D topological insulators, in particular those realised in HgTe/HgCdTe
and InAs/GaSb quantum wells. For example, recent observation of Josephson-junction interference
patterns in an S-HgTe/HgCdTe-S hybrid system has enabled extraction of \textit{g} factors for the
quantum-well charge carriers~\cite{Hart2015arXiv}. It would be interesting to use similar
techniques~\cite{Hart2014NPhys,Pribiag2015NNano} to measure the edge-state \textit{g} factors and
compare with our predictions. Furthermore, our results for the spin response in both the intrinsic and
doped regimes are informative for the design of, and interpretation of measured quantities for, dilute
magnetic phases in these systems~\cite{Novik2005PRB,Dietl2010NatMat}.

It would be interesting to extend our formalism to study the spin response in 3D topological-insulator
materials~\cite{Hasan2011AnnuRev}. In particular, as was the case in the 2D quantum-well-based TIs
considered in this work, the interplay of charge-carrier dynamics and the spinor character of extended
bulk states could be a source of rich variety in spin-related properties also in 3D, and the
contributions of the conducting surfaces are currently not understood.

\begin{acknowledgments}
E.~M.~H.\ acknowledges financial support from German Science Foundation (DFG) Grant No.\ 
HA 5893/4-1 within SPP 1666 and from the ENB graduate school "Topological Insulators".
This work was started while E.~M.~H.\ and U.~Z.\ visited the Kavli Institute for Theoretical
Physics at the University of Santa Barbara where their work was supported in part by the National
Science Foundation under Grant No.\ NSF PHY11-25915. The authors thank R.\ Winkler for
illuminating discussions about possible forms of relevant spin operators and useful comments
that have improved the manuscript.
\end{acknowledgments}


\begin{appendix}

\section{Intrinsic contribution to the spin susceptibility\label{sec:IntrinsicNum}}

\begin{figure}[t]
\includegraphics[width=0.9\columnwidth]{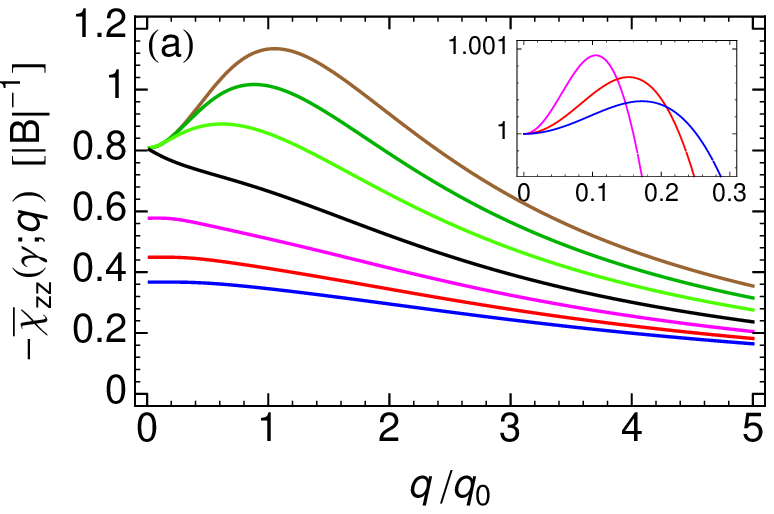}\\[-1cm]
\includegraphics[width=0.9\columnwidth]{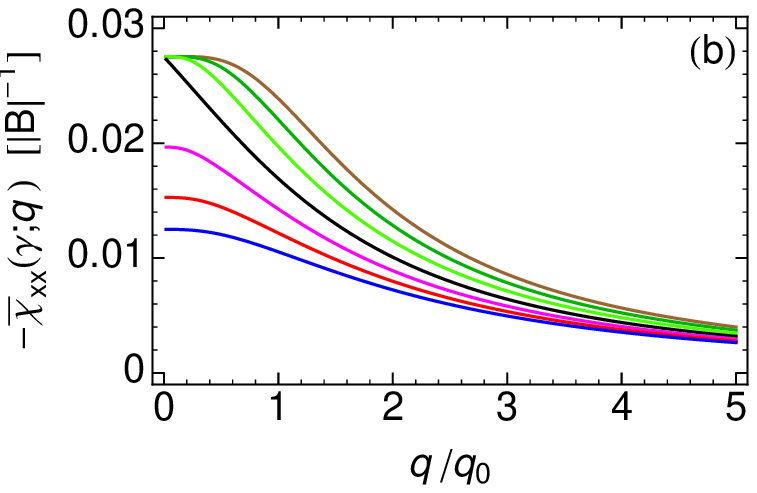}\\[-0.5cm]
\caption{\label{fig:Spinsusq}
Intrinsic spin susceptibilities (a) $\overline{\chi}_{zz} (\gamma;q)$ and (b) $\overline{\chi}_{xx} 
(\gamma;q)$ as a function of $q/q_0$ for various values of $\xi_{\text{M}}=(-0.3, -0.2, -0.1, 0, 0.1, 0.2,
0.3)$ (from top to bottom) with $\xi_{\text{D}}=-0.7$, $\mathcal{C}_{\text{LH}}=0.4$ and $\gamma=
-2.22$. The inset in (a) displays the normalized spin susceptibility $\overline{\chi}_{zz} (\gamma;q)/
\overline{\chi}_{zz} (\gamma;0)$ in the small $q/q_0$ region for $\xi_{\text{M}}=(0.1, 0.2,0.3)$, which
shows more clearly that $\overline{\chi}_{zz} (\gamma;q)>\overline{\chi}_{zz} (\gamma;0)$ in the
small-$q$ limit also in the normal regime.}
\end{figure}

The intrinsic contributions to the diagonal entries of the spin susceptibility (the only ones that are
non-zero for the BHZ model) are calculated by
\begin{eqnarray}\label{eq:Intrinsic}
\overline{\chi}_{jj}^{(\text{int})}(\gamma;\qq) &=&
-\sum_{s,s'\atop \delta=\pm 1}
\int\frac{d^2k}{(2\pi)^2} ~~\frac{\mathscr{W}_{jj(\kk,\kk+\qq,+,-)}^{(s,s')}(\gamma)~ n_{\text{F}}
\big(E^{(s)}_{\kk -}\big)}{E^{(s)}_{\kk +} - E^{(s')}_{\kk+\qq -}
+i\hbar\eta\delta}~, \nonumber \\
\end{eqnarray}
where in the zero-temperature limit the valence band is fully occupied, i.e., $n_{\text{F}}
\big(E^{(s)}_{\kk -}\big)=1$. Here and in the following we consider the growth-direction-averaged
case: $\overline{\chi}_{ij} (\gamma;\vek{q}) =  \int d z \int dz' ~ \chi_{ij}(\gamma;\vek{q}; z, z')
\Rightarrow\mathscr{W}_{jj(\kk,\kk+\qq,\alpha,\beta)}^{(s,s')}(\gamma)= \int dz~dz'~\mathscr{W}_{jj
(\kk,\kk+\qq,\alpha,\beta)}^{(s,s')}(\gamma;z,z')$. In Fig.~\ref{fig:Spinsusq}, we plot
$\overline{\chi}_{zz} (\gamma;q)$ [$\overline{\chi}_{xx} (\gamma;q)$] in panel (a) [(b)] as a function
of $q$ for various values of $\xi_{\text{M}}$. The inset in Fig.~\ref{fig:Spinsusq}(a) illustrates that
$\left.\frac{\partial^2 \overline{\chi}^{(\text{int})}_{zz}(\gamma;q)}{\partial q^2}\right|_{q=0}<0$
for all of the values of $\xi_{\text{M}}$ considered, which implies that the mean-field ferromagnetic
order for out-of-plane aligned magnetic impurity spins will generally be destroyed by spin-wave
(magnon) excitations~\cite{Simon2008PRB}.

\section{Bloembergen-Rowland behavior of the local intrinsic spin susceptibility
\label{sec:BloembergenRowland}}

\begin{figure}[t]
\includegraphics[width=0.8\columnwidth]{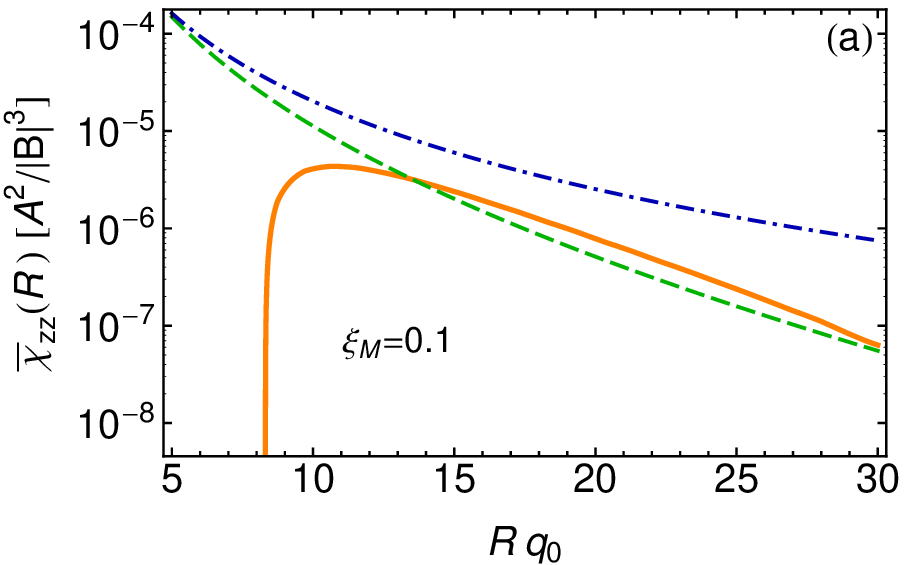}\\[0.5cm]
\includegraphics[width=0.8\columnwidth]{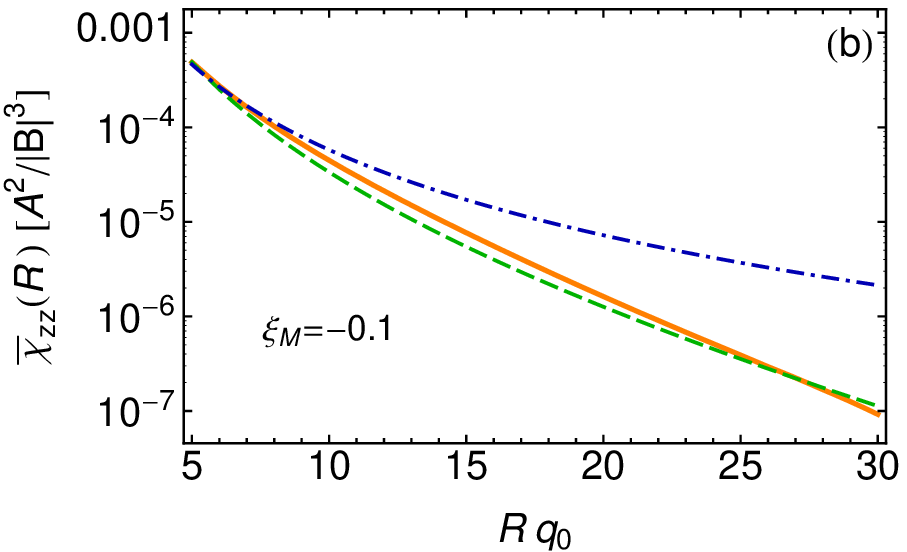}
\caption{\label{fig:BRchiR}
Local intrinsic spin susceptibilities $\overline{\chi}_{zz} (R)$ (solid orange curve) as a function of
$R q_0$ for (a) $\xi_{\text{M}}=0.1$ and (b) $\xi_{\text{M}}=-0.1$ with $\xi_{\text{D}}=-0.7$,
$\mathcal{C}_{\text{LH}}=0.4$ and $\gamma=-2.22$. The dashed green curve shows the modelled
behavior in Eq.~(\ref{eq:RKKYBRmodel}) with coefficients $c=1$ for (a) and $c=1.2$ for (b). For
comparison we also show the $R^{-3}$ decay for a gapless Dirac system represented by the
dot-dashed blue curve. The sharp drop of the orange curve in (a) is due to a sign change of
$\overline{\chi}_{zz} (R)$ at about $Rq_0\approx 8$.}
\end{figure}

Bloembergen and Rowland~\cite{Bloembergen1955PRB} found that the local RKKY interaction of 
gapped systems becomes short-ranged, i.e., is exponentially suppressed by the band-gap. The
functional dependence of the local spin susceptibility on the distance for the gapped Dirac system at
hand can thus be modelled by
\begin{eqnarray}\label{eq:RKKYBRmodel}
\overline{\chi}_{jj}(R)\sim \frac{\ee^{-c\frac{R}{\lambda_{\text{C}}}}}{R^3}~,
\end{eqnarray}
where $\lambda^{-1}_{\text{C}}\equiv|\xi_{\text{M}}|q_0$ is the inverse of the Compton wavelength of 
the system and $c\sim\mathcal{O}(1)$ is a numerical coefficient that can depend on the distance itself.
To illustrate this behavior for the HgTe quantum well system, we plot $\overline{\chi}_{zz}(R)$ in
Fig.~\ref{fig:BRchiR} as a function of $Rq_0$ for $\xi_{\text{M}}=0.1$ [$\xi_{\text{M}}=-0.1$] in panel
(a) [(b)] together with both the line shape expected for 2D massless-Dirac particles and the
Bloembergen-Rowland result.

\begin{figure*}[t]
\includegraphics[width=0.9\columnwidth]{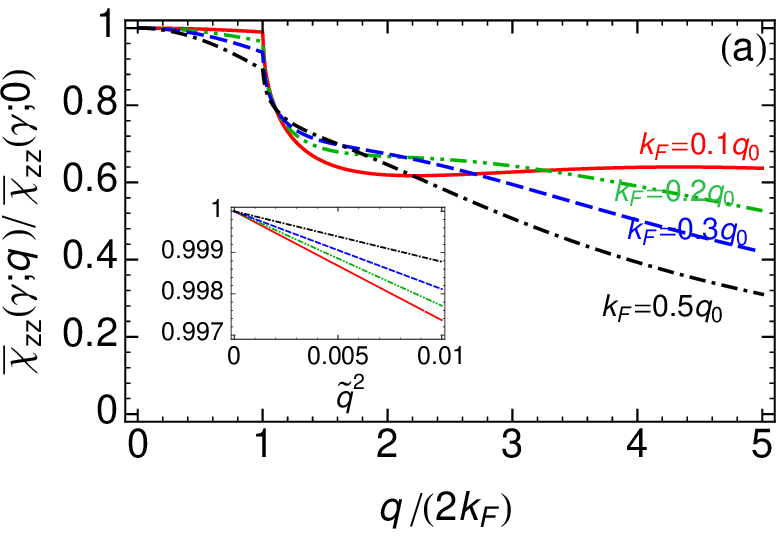}
\includegraphics[width=0.9\columnwidth]{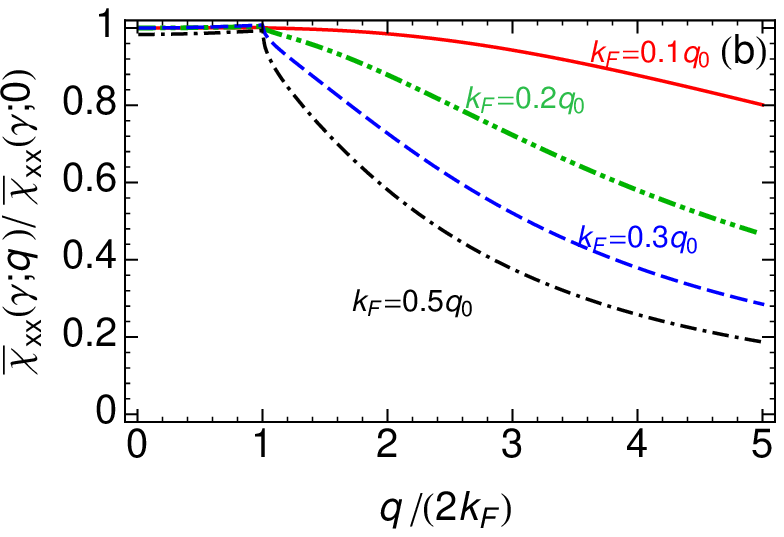}
\vskip-0.5cm
\includegraphics[width=0.9\columnwidth]{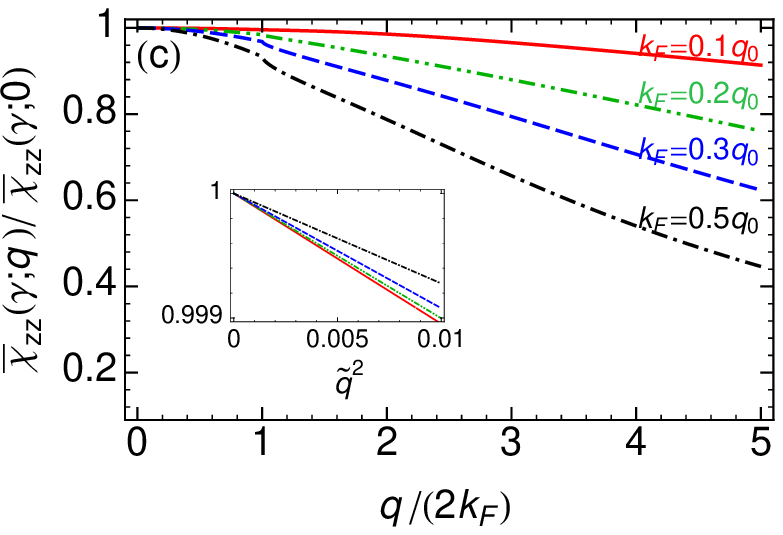}
\includegraphics[width=0.9\columnwidth]{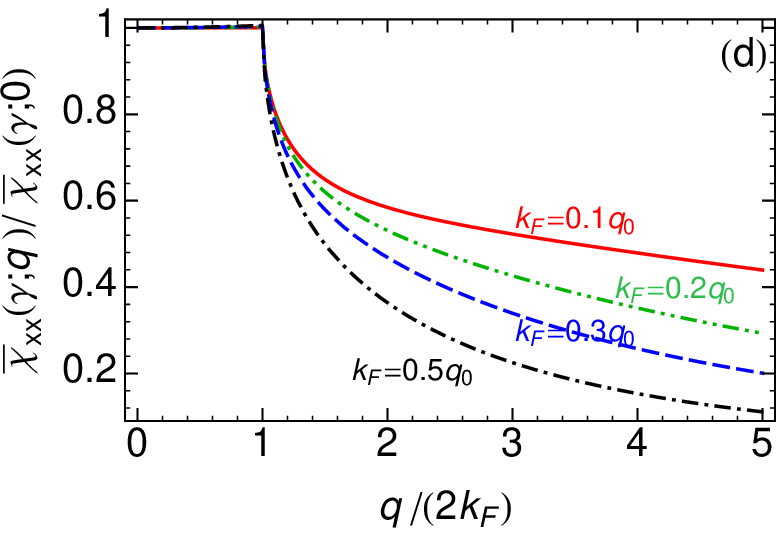}
\caption{\label{fig:SpinSusdop}
Normalized spin susceptibilities $\overline{\chi}_{zz}(\gamma;q)$ (a,c) and $\overline{\chi}_{xx}
(\gamma;q)$ (b,d) as a function of $q/(2k_{\text{F}})$ for various levels of doping. (a,b) are the
results for $\xi_{\text{M}}=-0.2$ and (c,d) for $\xi_{\text{M}}=0.2$, with $\xi_{\text{D}}=-0.7$,
$\mathcal{C}_{\text{LH}}=0.4$ and $\gamma=-2.22$. The insets of (a,c) show the quadratic
$\tilde q=q/q_0$ dependence close to $q=0$.}
\end{figure*}

\section{Electron doped contribution to the spin susceptibility\label{sec:DopedNum}}

For the case where the Fermi energy is above the conduction energy band edge, i.e., $\mu>|M|$,
the spin susceptibility receives contributions due to electron doping given by
\begin{eqnarray}\label{eq:Eldopedgeneric}
\overline{\chi}_{jj}^{(\text{dop})}(\gamma;\qq) &=& \sum_{s,s'\atop \delta=\pm 1}
\int\frac{d^2k}{(2\pi)^2}~n_{\text{F}}\big(E^{(s)}_{\kk +}\big) \nonumber \\[2mm]
&& \hspace{-1.3cm} \times\left[
\frac{\mathscr{W}_{jj(\kk,\kk+\qq,+,+)}^{(s,s')}(\gamma)}{E^{(s)}_{\kk +}-E^{(s')}_{\kk
+\qq +} +i\hbar\eta\delta}+\frac{\mathscr{W}_{jj(\kk,\kk+\qq,+,-)}^{(s,s')}(\gamma)}{E^{(s)}_{\kk +}
-E^{(s')}_{\kk+\qq -}+i\hbar\eta\delta}\right], \nonumber\\[2mm]
\end{eqnarray}
where for zero temperature $n_{\text{F}}\big(E^{(s)}_{\kk +}\big)=\Theta\big(k_{\text{F}}-|\kk|\big)$,
with $k_{\text{F}}$ being the Fermi wave vector associated with the conduction band. The complete
spin susceptibility in the doped case is therefore given by
\begin{eqnarray}\label{eq:Elchi}
\overline{\chi}_{jj}(\gamma;\qq)=\overline{\chi}_{jj}^{(\text{int})}(\gamma;\qq)+
\overline{\chi}_{jj}^{(\text{dop})}(\gamma;\qq)~.
\end{eqnarray}
We show the line shape of $\overline{\chi}_{zz}(\gamma;q)$ and $\overline{\chi}_{xx}(\gamma;q)$ in
Fig.~\ref{fig:SpinSusdop} for $\xi_{\text{M}}=\pm0.2$ and various choices of doping. The insets of
Fig.~\ref{fig:SpinSusdop}(a,c) show the small $q$ dependence of $\overline{\chi}_{zz}(\gamma;q)$,
indicating the stability of the out-of-plane Ising-type ferromagnetism with respect to Magnon
excitations. 

\section{Combining bulk and edge contributions to the spin susceptibility\label{sec:compbulkedge}}

In order to compare bulk and edge contributions to the spin susceptibility we begin by considering the
real-space spin-response function
\begin{equation}
\chi_{ij}(\vek{r}, \vek{r^\prime}) = -\frac{i}{\hbar} \int_0^\infty d t \,\,\, \ee^{-\eta t}\,\, \langle\left[
S_i(\vek{r}, t)\, , \, S_j(\vek{r^\prime}, 0)\right]\rangle\,\, ,
\end{equation}
where $S_i(\vek{r}) = \Psi^\dagger(\vek{r})\hat S_i \Psi(\vek{r})$ are spin-density operators.
In our system of interest, electrons shall be confined to move freely in $d<3$ dimensions,
with system size $L$ in all of these free directions. The position vector shall be split up
into a part $\vek{R}$ comprising the coordinate directions in which the motion is free and a
part $\vek{\varrho}$ in whose coordinates motion is confined; $\vek{r} = (\vek{R}, \vek{\varrho})$.
The second-quantised electron operator in real-space representation can be written as
\begin{equation}
\Psi(\vek{r}) = \sum_{\vek{k},\alpha,s} \frac{\ee^{i\vek{k}\cdot\vek{R}}}{\sqrt{L^d}}\,\,
\psi^{(s)}_{\vek{k}\alpha}(\vek{\varrho}) \,\, c^{(s)}_{\vek{k}\alpha}
\end{equation}
with normalized spinor bound-state wave functions $\xi_{n\vek{k}}(\vek{\varrho})$. A straightforward
calculation yields
\begin{subequations}
\begin{equation}
\chi_{ij}(\vek{r}, \vek{r^\prime}) = \frac{1}{L^d}\, \sum_\vek{q} \, \ee^{i\vek{q}\cdot(\vek{R}
-\vek{R^\prime})}\,\, \chi_{ij}^{(d\text{D})}(\vek{q}; \vek{\varrho}, \vek{\varrho^\prime})\quad ,
\end{equation}
with the $\vek{q}$-dependent spin susceptibility of the $d$-dimensional ($d$D) system given by
\begin{eqnarray}
\chi_{ij}^{(d\text{D})}(\vek{q}; \vek{\varrho}, \vek{\varrho^\prime}) &=& \sum_{\alpha,\beta,s,s'} \,
\frac{1}{L^d}\, \sum_\vek{k}\mathscr{W}_{ij(\kk,\kk+\qq,\alpha,\beta)}^{(s,s')}(\vek{\varrho},
\vek{\varrho^\prime}) \nonumber \\ && \hspace{1cm} \times\, \frac{n_{\text F}(E^{(s)}_{\kk\alpha})
-n_{\text F}(E^{(s')}_{\kk+\qq\beta})}{E^{(s)}_{\kk\alpha}-E^{(s')}_{\kk+\qq\beta} +i\hbar\eta} \, . \qquad
\end{eqnarray}
\end{subequations}

We are now interested in the homogeneous part of the spin response defined as
\begin{subequations}
\begin{equation}
\Upsilon_{ij} = \int d^3 r\,\, \int d^3 r^\prime \,\,\,\, \chi_{ij}(\vek{r}, \vek{r^\prime}) 
\equiv L^d\, \overline{\chi}_{ij}^{(d\text{D})}(\vek{q}=0) \quad ,
\end{equation}
where
\begin{equation}
\overline{\chi}_{ij}^{(d\text{D})}(\vek{q}) = \int d^{3-d}\varrho \,\, \int d^{3-d}\varrho^\prime \,\,\,\, 
\chi_{ij}^{(d\text{D})}(\vek{q}; \vek{\varrho}, \vek{\varrho^\prime}) \quad .
\end{equation}
\end{subequations}
In the situation where both 2D bulk and 1D edge states are present, we therefore find
\begin{equation}
\Upsilon_{ij} = L^2 \left[ \overline{\chi}_{ij}^{(\text{2D})}(\vek{q}=0) + \frac{1}{L}\,
\overline{\chi}_{ij}^{(\text{1D})}(q=0) \right] \quad .
\end{equation}
Thus $L^{-2} \Upsilon$ is the well-defined quantity in the thermodynamic limit, and only those
edge-related terms that scale with $L$ will contribute to it.

\section{Spin susceptibility of edge states\label{sec:spinsusedge}}

\subsection{Spin susceptibility of edge states: Omitting finite size effects}

Following~\cite{Qi2011RMP} (see also~\cite{Zhou2008PRL}), the dispersions for the edge states, 
using open boundary conditions for a confinement along the $x$ direction and assuming $D=0$, is
given by
\begin{eqnarray}\label{eq:edgedispersion}
E^{(s)}_{k}=sA k~,
\end{eqnarray}
where $k\equiv k_y$.
The associated spinor wave functions are
\begin{eqnarray}\label{eq:edgestatespinors}
\eta^{(s)}(\vek{\varrho})&=&\sum_{l=1}^2\varphi^{(s)}_{0,l}(x)~\psi^{(s)}_{0l}(z)~,
\end{eqnarray}
where $\varphi^{(s)}_{0}(x)=C(\ee^{\lambda_1x}-\ee^{\lambda_2x})\phi_{-s}$, with $\phi_\pm^T=(1,
\pm i)$ and $\lambda_{1,2}=-\frac{q_0}{2}(1\pm\sqrt{1-4|\xi_{\text{M}}}|)$, since $M<0$. To simplify
matters, we have assumed particle-hole symmetry ($D=0$) and the system length $L\to \infty$, i.e.,
the gap in the dispersions of the edge states is negligible. Note that the spinors in
Eq.~(\ref{eq:edgestatespinors}) are independent of the wave vector component along the $y$
direction.

The spin susceptibility of the edge states is calculated by
\begin{eqnarray}\label{eq:spinsusedge}
\chi^{\text{(edge)}}_{jj}(\gamma;q;\vek{\varrho},\vek{\varrho^\prime})&=&\sum_{s,s'}\int \frac{dk}{2\pi}
~\mathscr{W}_{jj}^{(s,s')}(\gamma;\vek{\varrho},\vek{\varrho^\prime})\nonumber\\[1mm]
{}&&\times \frac{n_{\text{F}}[E^{(s)}_{k}]-n_{\text{F}}[E^{(s')}_{k+q}]}{E^{(s)}_{k}-E^{(s')}_{k+q}+i\hbar
\eta}~,
\end{eqnarray}
where 
\begin{eqnarray}
\mathscr{W}_{jj}^{(s,s')}(\gamma;\vek{\varrho},\vek{\varrho^\prime})&=&[\eta^{(s)}(\vek{\varrho})
]^{\dagger} \cdot\hat S_j(\gamma) \cdot \eta^{(s')}(\vek{\varrho}) \nonumber\\[1mm] {}&& \times 
[\eta^{(s')}(\vek{\varrho^\prime})]^\dagger\cdot\hat S_j(\gamma) \cdot\eta^{(s)}(\vek{\varrho^\prime}).
\end{eqnarray}
With an averaging over the coordinates along the confined directions we have
\begin{eqnarray}
\int d\vek{\varrho}~d\vek{\varrho^\prime}~\mathscr{W}_{zz}^{(s,s')}(\gamma;\vek{\varrho},
\vek{\varrho^\prime})&=& \frac{\Xi^2_z(\gamma)}{16}\begin{pmatrix} 
1 & 0\\
0 & 1
\end{pmatrix}~, \\[2mm]
\int d\vek{\varrho}~d\vek{\varrho^\prime}~\mathscr{W}_{xx}^{(s,s')}(\gamma;\vek{\varrho},
\vek{\varrho^\prime})&=& \frac{\Xi^2_x(\gamma)}{16}\begin{pmatrix} 
0 & 1\\
1 & 0
\end{pmatrix}~,
\end{eqnarray}
with $\Xi_z(\gamma)=1+3\gamma+(\gamma-1)\mathcal{C}_{\text{LH}}$ and $\Xi_x
(\gamma)=1+(2\gamma-1)\mathcal{C}_{\text{LH}}\equiv {\mathcal C}_x(\gamma)$.
Thus $\chi^{(\text{edge})}_{zz}(\gamma;q)\propto\chi^{(\text{edge})}_{0}(q)$, where
$\chi^{(\text{edge})}_{0}(q)$ is the Lindhard function associated with the edge states.
Explicit calculation of (\ref{eq:spinsusedge}) yields
\begin{subequations}\label{eq:susedgeelec}
\begin{eqnarray}\label{eq:susedgezz}
\chi_{zz}^{(\text{edge})}(\gamma;q) &=& - \frac{\Xi^2_z(\gamma)}{16\pi A} \equiv \frac{\Xi^2_z 
(\gamma)}{16}\, \chi^{(\text{edge})}_{0}(q)~, \\ \label{eq:susedgexx}
\chi_{xx}^{(\text{edge})}(\gamma;q)&=&\frac{\Xi^2_x(\gamma)}{32\pi A}\ln\left(\frac{|q^2 - 4 
k^2_{\text{F}}|}{4\Lambda^2-q^2}\right)~,
\end{eqnarray}
\end{subequations}
where $\Lambda$ is a large-wave-vector cut-off. Due to the special energy dispersion 
Eq.~(\ref{eq:edgedispersion}), $\chi_{zz}^{(\text{edge})}(\gamma;q)$ is a constant (independent of 
$k_{\text{F}}$ and $q$).
Clearly, for the hole doped case ($\mu<0$), the same result as in Eq.~(\ref{eq:susedgeelec}) is 
obtained due to the assumed particle-hole symmetry.

\subsection{Spin susceptibility of edge states: Finite size effects included}

Following~\cite{Zhou2008PRL}, we now take into account the finite size of the system. As a result, 
edge states at the two sides that have the same spin can couple which results in a gapped spectrum 
of their energy dispersions.We assume the system size $L$ to be large and use an approximation of 
the wave functions in~\cite{Zhou2008PRL}. Taking into account the various degrees of freedom of 
both system sides, we modify the approach of~\cite{Qi2011RMP}, where the solutions for the wave 
functions read
\begin{eqnarray}\label{eq:Finitsizewavefun}
\varphi^{(s)}_{\nu}(x)=\frac{1}{\sqrt{C}}\left(\ee^{\lambda_1(\frac{L}{2}-s\nu x)}-\ee^{\lambda_2
(\frac{L}{2}-s\nu x)}\right)\phi_{-s\nu}~,\nonumber\\
\end{eqnarray}
where $C$ is a normalization constant and $\nu=\pm$ denotes the right- and left-mover. For non-zero 
wave vector $k$, we consider the matrix
\begin{eqnarray}\label{eq:RLmover}
{\mathcal{H}_{\text{RL}}}=
\begin{pmatrix} 
A k & \Delta \\[2mm]
\Delta  & -A k 
\end{pmatrix}~
\end{eqnarray}%
in the basis of right- and left-mover, where $\Delta$ is the induced gap which is a function of the 
system parameters~\cite{Zhou2008PRL}. The energy dispersions of (\ref{eq:RLmover}) are 
\begin{equation}
E^{(s)}_{k\tau}=\tau\sqrt{(A k)^2+\Delta^2}~,
\end{equation}
where $\tau=\pm$ labels the (spin-degenerate) conduction and valence bands, respectively.
The associated eigenstates are
\begin{equation}
a^{(s)}_{k\tau}=\begin{pmatrix} 
\sqrt{\frac{E^{(s)}_{k\tau}+Ak}{2E^{(s)}_{k\tau}}}  \\[3mm]
\tau \sqrt{\frac{E^{(s)}_{k\tau}-Ak}{2E^{(s)}_{k\tau}}}
\end{pmatrix}~.
\end{equation}
With this information, the spinors in Eq.~(\ref{eq:edgestatespinors}) are modified to
\begin{eqnarray}
\eta^{(s)}_{k\tau}(\vek{\varrho})&=&\sum_{l=1}^2\sum_{\nu}a^{(s)}_{k\tau,\nu}~\varphi^{(s)}_{\nu,l}(x)~
\psi^{(s)}_{0l}(z)~,
\end{eqnarray}
Thus, for the present case, the spin susceptibility of the edge states is
\begin{eqnarray}\label{eq:spinsusedgefinitsize}
\chi^{\text{(edge)}}_{jj}(\gamma;q;\vek{\varrho},\vek{\varrho^\prime})&=&
\sum_{\substack{s,s' \\
\tau,\tau'}}
\int \frac{dk}{2\pi}~\mathscr{W}_{jj(k,k+q,\tau,\tau')}^{(s,s')}(\gamma;\vek{\varrho},\vek{\varrho^\prime})
\nonumber\\[1mm] {}&&\times\frac{n_{\text{F}}[E^{(s)}_{k\tau}]-n_{\text{F}}[E^{(s')}_{k+q
\tau'}]}{E^{(s)}_{k\tau}-E^{(s')}_{k+q\tau'}+i\hbar\eta}~,
\end{eqnarray}
where 
\begin{eqnarray}
\mathscr{W}_{jj(k,k+q,\tau,\tau')}^{(s,s')}&(\gamma;\vek{\varrho},\vek{\varrho^\prime})&=[\eta^{(s)}_{k
\tau}(\vek{\varrho})]^\dagger\cdot\hat S_j(\gamma)\cdot \eta^{(s')}_{k+q\tau'}(\vek{\varrho})\nonumber
\\[1mm]{}&& \times [\eta^{(s')}_{k+q\tau'}(\vek{\varrho^\prime})]^\dagger\cdot\hat S_j(\gamma)\cdot
\eta^{(s)}_{k\tau}(\vek{\varrho^\prime})~.\nonumber\\[2mm]
\end{eqnarray}
Averaging over the coordinates along the confined directions, we obtain for the
overlap factor of the Lindhard function
\begin{eqnarray}\label{eq:Overap0edge}
&&\int d\vek{\varrho}~d\vek{\varrho^\prime}~\mathscr{W}_{0(k,k+q,\tau,\tau')}^{(s,s')}(\vek{\varrho},
\vek{\varrho^\prime})=\frac{\delta_{ss'}}{4}\nonumber\\ &&{}\times
\begin{cases}
           (a_k a_{k+q}+b_k b_{k+q})^2       &\quad (\tau=\tau') \\[3mm]
           (a_k b_{k+q}-b_k a_{k+q})^2     &\quad (\tau\neq\tau')
\end{cases}
\end{eqnarray}
where $a_k\equiv \sqrt{1-\frac{A k}{\sqrt{(A k)^2+\Delta^2}}}$, $b_k\equiv \sqrt{1+\frac{A k}{\sqrt{(A 
k)^2+\Delta^2}}}$. The result for the overlap factors of the spin susceptibilities is
\begin{eqnarray}\label{eq:Overap0edgezz}
&&\int d\vek{\varrho}~d\vek{\varrho^\prime}~\mathscr{W}_{zz(k,k+q,\tau,\tau')}^{(s,s')}(\gamma;
\vek{\varrho},\vek{\varrho^\prime})=\frac{\delta_{ss'}}{64}\nonumber\\[2mm] &&{}\times
\begin{cases}
           [(a_k a_{k+q}+b_k b_{k+q})~\Xi_z(\gamma) &\quad (\tau=\tau')\\[1mm]
           +\tau (a_k b_{k+q}+b_k a_{k+q})~\mathcal{C}_z(\gamma)~N L \Delta]^2        \\[3mm]
           [(b_k b_{k+q}-a_k a_{k+q})~\mathcal{C}_z(\gamma)~N L \Delta&\quad (\tau\neq\tau')\\[1mm]
           +\tau(a_k b_{k+q}-b_k a_{k+q})~\Xi_z(\gamma)]^2      
\end{cases}
\end{eqnarray}
and
\begin{eqnarray}\label{eq:Overap0edgexx}
&&\int d\vek{\varrho}~d\vek{\varrho^\prime}~\mathscr{W}_{xx(k,k+q,\tau,\tau')}^{(s,s')}(\gamma;
\vek{\varrho},\vek{\varrho^\prime}) =\frac{\Xi^2_x(\gamma)}{64}(1-\delta_{ss'})\nonumber\\[2mm]
&&{}\times
\begin{cases}
[a_k b_{k+q}+ b_k a_{k+q}&\quad (\tau=\tau')\\[1mm]
+\tau(a_k a_{k+q}+ b_k b_{k+q}) N L \Delta ]^2\\[3mm]
[a_k a_{k+q}- b_k b_{k+q}&\quad (\tau\neq \tau')\\[1mm]
-\tau(a_k b_{k+q}- b_k a_{k+q}) N L \Delta ]^2\\[3mm]
\end{cases}
\end{eqnarray}
In obtaining (\ref{eq:Overap0edge})-(\ref{eq:Overap0edgexx}) we have used 
\begin{equation}
\int~dx\prod_\nu \left(\ee^{\lambda_1(\frac{L}{2}-\nu x)}-\ee^{\lambda_2(\frac{L}{2}-\nu x)}\right)
\approx L ~\ee^{\lambda_2 L}~,
\end{equation} 
the functional $L$-dependence of $\Delta \approx F~ \text{exp}(\lambda_2 L)$~\cite{Zhou2008PRL}, 
where $F/E_0=4|\xi_{\text{M}}|/\sqrt{1-4|\xi_{\text{M}}|}$, and we have defined $N\equiv (F C)^{-1}$.

\subsubsection{Intrinsic contribution to the spin susceptibilities of the edge states in the limit $q\to 0$}

Calculating the intrinsic contribution to the spin susceptibilities of the edge states in the
long-wavelength limit ($q\to 0$), we obtain
\begin{subequations}\label{eq:intrinsicq0edge1}
\begin{eqnarray}\label{eq:intrinsicq0edge1zz}
\chi^{(\text{int,e})}_{zz}(\gamma;0) =(N L \Delta)^2\frac{g_{\text{s}}\mathcal{C}^2_z(\gamma)}{16\pi A}
\left[1-\ln\left(\frac{2\Lambda}{\tilde\Delta}\right)\right]~,\nonumber\\
\end{eqnarray}
\begin{eqnarray}\label{eq:intrinsicq0edge1xx}
\chi^{(\text{int,e})}_{xx}(\gamma;0)=\frac{g_{\text{s}}\Xi^2_x(\gamma)}{16\pi A}\left[1-\ln
\left(\frac{2\Lambda}{\tilde\Delta}\right)\right]~,
\end{eqnarray}
\end{subequations}
where $\tilde\Delta\equiv \Delta/A$. We note that the intrinsic contribution to the Lindhard function 
vanishes in the limit $q\to 0$, which can be inferred from (\ref{eq:Overap0edge}). To compare this
result with the one of the intrinsic bulk contribution, we divide (\ref{eq:intrinsicq0edge1zz}) and 
(\ref{eq:intrinsicq0edge1xx}) by the length $L$ and let $L$ go to infinity (see
Sec.~\ref{sec:compbulkedge}). Thus, we obtain
\begin{subequations}\label{eq:Edgeq0Intrinsic}
\begin{eqnarray}
\lim_{L\to \infty}\frac{\chi^{(\text{int,e})}_{zz}(\gamma;0)}{L}=0~,
\end{eqnarray}
\begin{eqnarray}\label{eq:Edgeq0Intrinsicxx}
\lim_{L\to \infty}\frac{\chi^{(\text{int,e})}_{xx}(\gamma;0)}{L}=-\frac{g_{\text{s}}\Xi^2_x(\gamma)}{32\pi
|B|}\left(1-\sqrt{1-4|\xi_{\text{M}}|}\right)~.\nonumber\\
\end{eqnarray}
\end{subequations}
Therefore, the edge states give a contribution to the total susceptibility only for the in-plane component 
and its sign equals that of the bulk contribution.

\subsubsection{Electron doped contribution to the spin susceptibilities of the edge states in the limit
$q\to 0$}

Next we include the contributions due to doping to the spin susceptibilities. The interband contributions 
read (for $A k_{\text{F}}\gg\Delta$)
\begin{subequations}\label{eq:dopedq0edgeinter}
\begin{eqnarray}\label{eq:dopedq0edgeinterzz}
\chi^{(\text{inter,e})}_{zz}(\gamma;0)
=-(NL\Delta)^2\frac{g_{\text{s}}\mathcal{C}_z^2(\gamma)}{16\pi A}\left[1-\ln\left(
\frac{2k_{\text{F}}}{\tilde\Delta}\right)\right],\nonumber\\
\end{eqnarray}
\begin{eqnarray}\label{eq:dopedq0edgeinterxx}
\chi^{(\text{inter,e})}_{xx}(\gamma;0)
=-\frac{g_{\text{s}}\Xi^2_x(\gamma)}{16\pi A}\left[1-\ln\left(\frac{2k_{\text{F}}}{\tilde\Delta}\right)\right],
\end{eqnarray}
\end{subequations}
while the intraband contributions are
\begin{subequations}\label{eq:dopedq0edgeintra}
\begin{eqnarray}\label{eq:dopedq0edgeintrazz}
\chi^{(\text{intra,e})}_{zz}(\gamma;0)
=-\frac{g_{\text{s}}\Xi^2_z(\gamma)}{16\pi A}~,
\end{eqnarray}
\begin{eqnarray}\label{eq:dopedq0edgeintraxx}
\chi^{(\text{intra,e})}_{xx}(\gamma;0)=-(NL\Delta)^2~\frac{g_{\text{s}}\Xi^2_x(\gamma)}{16\pi A}~,
\end{eqnarray}
\end{subequations}
which is consistent with the finding that $\chi^{(\text{intra,e})}_{zz}(\gamma;q)$
[$\chi^{(\text{intra,e})}_{xx}(\gamma;q)$] are important [unimportant] for $\Delta\to 0$, whereas it is
the other way around for the interband contributions.
Considering 
\begin{subequations}
\begin{eqnarray}
\lim_{L\to \infty}&&\frac{[\chi^{(\text{intra,e})}_{zz}(\gamma;0)+
\chi^{(\text{inter,e})}_{zz}(\gamma;0)]}{L}=0~,
\end{eqnarray}
\begin{eqnarray}
\lim_{L\to \infty}&&\frac{[\chi^{(\text{intra,e})}_{xx}(\gamma;0)+\chi^{(\text{inter,e})}_{xx}
(\gamma;0)]}{L}=\nonumber\\[2mm] &&{}\hskip1.4cm\frac{g_{\text{s}}\Xi^2_x(\gamma)}{32\pi|B|}
\left(1-\sqrt{1-4|\xi_{\text{M}}|}\right),
\end{eqnarray}
\end{subequations}
we find that this is the same contribution as the intrinsic contribution, Eq.~(\ref{eq:Edgeq0Intrinsicxx}), 
which has however the opposite sign. Thus, the sum of doped and intrinsic contributions of the edge
states vanishes in the large $L$ limit. As a consistency check, we verify for $q=0$ that the sum of
Eqs.~(\ref{eq:intrinsicq0edge1}), (\ref{eq:dopedq0edgeinter}) and (\ref{eq:dopedq0edgeintra}) yields
Eqs.~(\ref{eq:susedgezz}) and (\ref{eq:susedgexx}) (multiplied by $g_{\text{s}}$) in the limit $\Delta
\to 0$. 
%


\section{Effects of structural inversion asymmetry}

Here we demonstrate that the basic features of the intrinsic spin susceptibilities given in 
Eq.~(\ref{eq:Intrinsicq0}) are robust even when effects due to structural inversion asymmetry (SIA)
are included. By taking into account the influence of a perpendicular electric field $\mathcal{E}_z$, it
has been shown in Ref.~\cite{Rothe2010NJP} that the BHZ Hamiltonian is supplemented by entries
that mix the spin-up and spin-down components of the BHZ basis states. The leading contribution
due to SIA is linear in the wave vector and given by
\begin{equation}\label{eq:effSIA}
\mathscr{H}_{\text{R}}=
\begin{pmatrix} 
0 & 0 & -i R_0k_- & 0  \\[2mm]
0& 0& 0 & 0  \\[2mm]
i R_0k_+ & 0 & 0 & 0 \\[2mm]
0 & 0 & 0& 0
\end{pmatrix}~,
\end{equation}
where $k_\pm=k_x\pm i k_y$. The necessity to avoid dielectric breakdown provides an upper limit 
for the electric-field magnitude through the condition $|e\mathcal{E}_z| d < 2|M|$. Defining
the SIA-related dimensionless parameter $\xi_{\text{R}} \equiv R_0/A$, this condition translates into
$|\xi_{\text{R}}| < 16.1\, |\xi_{\text{M}}|/(d\, [\mbox{nm}])\approx 0.12$ for a typical
heterostructure~\cite{Rothe2010NJP}. Thus $\xi_R$ is generally a small parameter, and a perturbative
treatment for SIA effects is appropriate. To lowest order in $\xi_{\text{R}}$, the intrinsic spin susceptibility
in the limit $q\to 0$ is found as
\begin{widetext}
\begin{subequations}\label{eq:Intrinsicjjq0R}
\begin{eqnarray}\label{eq:Intrinsicxxq0R}
\overline{\chi}^{(\text{int})}_{xx}(\gamma; \qq = 0) &=& -\frac{\mathcal{C}^2_{x}(\gamma)}{16\pi |B|}
\frac{1}{1+4\xi_{\text{M}}\Theta(\xi_{\text{M}})} \, \left(1+\xi_{\text{R}}^2~\frac{8\xi_{\text{M}}
\Theta(\xi_{\text{M}})}{3[1+4\xi_{\text{M}}\Theta(\xi_{\text{M}})]^2}\right)~, \\ \label{eq:Intrinsiczzq0R}
\overline{\chi}^{(\text{int})}_{zz}(\gamma; \qq = 0) &=& -\frac{\mathcal{C}^2_{z}(\gamma)}{16\pi |B|}
\frac{1}{1+4\xi_{\text{M}}\Theta(\xi_{\text{M}})}\, \left(1 +\xi_{\text{R}}^2~\frac{2\{9\gamma^2+2 
\xi_{\text{M}} \Theta(\xi_{\text{M}})[\mathcal{C}^2_{z}(\gamma)+18\gamma^2]\}}{3\mathcal{C}^2_{z}
(\gamma)[1+4\xi_{\text{M}}\Theta(\xi_{\text{M}})]^2}\right)~.
\end{eqnarray}
\end{subequations}
\end{widetext}
Thus the lowest-order SIA corrections to $\overline{\chi}^{(\text{int})}_{jj}(\gamma; \qq = 0)$ are
quadratic in the small parameter $\xi_{\text{R}}$. This means that the result given in
Eq.~(\ref{eq:Intrinsicjjq0R}) represents already an excellent approximation. Interestingly,
$\overline{\chi}^{(\text{int})}_{xx}(\gamma; \qq = 0)$ turns out to not be modified by SIA contributions
in the inverted regime. We find that this remains true even when higher-order corrections in
$\xi_{\text{R}}$ are considered. In contrast, $\overline{\chi}^{(\text{int})}_{zz}(\gamma; \qq = 0)$ in
Eq.~(\ref{eq:Intrinsiczzq0R}) has finite SIA corrections in the inverted regime given by $6\gamma^2
\xi_{\text{R}}^2/\mathcal{C}^2_z(\gamma)$. In our case where $\gamma=-2.22$ this amounts to a
relative change that is about 1\%. Also in the normal regime, SIA contributions to
$\overline{\chi}^{(\text{int})}_{jj}(\gamma; \qq = 0)$ are at most of relative magnitude 1\%. Thus we
conclude that the spin susceptibilities in Eq.~(\ref{eq:Intrinsicq0}) generally receive only very small
corrections when SIA terms are included in the BHZ Hamiltonian. 

\end{appendix}


%

\end{document}